\documentclass[useAMS,usenatbib]{mn2e}

\usepackage{graphicx}
\usepackage{amsmath}
\usepackage{amssymb}
\usepackage{color}
\usepackage{subfig}

\usepackage{hyperref}
\hypersetup{colorlinks=true,linkcolor=blue,citecolor=blue,filecolor=blue,urlcolor=blue}

\newcommand\msun{\, \rm M_\odot}
\newcommand\rsun{\, \rm R_\odot}
\newcommand\kms{\, \rm km\,s^{-1}}
\newcommand\pc{{\, \rm pc}}
\newcommand\kpc{{\, \rm kpc}}
\newcommand\yr{{\, \rm yr}}

\newcommand\au{{\, \rm AU}}

\newcommand\rmin{{r_{\rm min}}}

\title[HVSs from star clusters hosting an IMBH]{Hypervelocity stars from star clusters hosting Intermediate-Mass Black Holes}
\author[G. Fragione and A. Gualandris]{Giacomo Fragione$^{1}$\thanks{E-mail: giacomo.fragione@mail.huji.ac.il} and Alessia Gualandris$^{2}$\\
$^{1}$Racah Institute for Physics, The Hebrew University, Jerusalem 91904, Israel\\
$^{2}$Department of Physics, University of Surrey, Guildford GU2 7XH, United Kingdom}

\begin{document}

\maketitle

\begin{abstract}
Hypervelocity stars (HVSs) represent a unique population of stars in the Galaxy reflecting properties of the whole Galactic potential. Determining their origin is of fundamental importance to constrain the shape and mass of the dark halo. The leading scenario for the ejection of HVSs is an encounter with the supermassive black hole in the Galactic Centre. However, new proper motions from the \textit{Gaia} mission indicate that only the fastest HVSs can be traced back to the Galactic centre and the remaining stars originate in the disc or halo. In this paper, we study HVSs generated by encounters of stellar binaries with an intermediate-mass black hole (IMBH) in the core of a star cluster. For the first time, we model the effect of the cluster orbit in the Galactic potential on the observable properties of the ejected population. HVSs generated by this mechanism do not travel on radial orbits consistent with a Galactic centre origin, but rather point back to their parent cluster, thus providing observational evidence for the presence of an IMBH. We also model the ejection of high-velocity stars from the Galactic population of globular clusters, assuming that they all contain an IMBH, including the effects of the cluster's orbit and propagation of the star in the Galactic potential up to detection. We find that high-velocity stars ejected by IMBHs have distinctive distributions in velocity, Galactocentric distance and Galactic latitude, which can be used to distinguish them from runaway stars and stars ejected from the Galactic Centre.
\end{abstract}


\section{Introduction}

The European Space Agency mission \textit{Gaia}\footnote{http://sci.esa.int/gaia/} has revolutionized astrometry, providing positions, parallaxes and proper motions for more than $1.3$ billion stars in its second data release (Gaia DR2) \citep{gaia18}. It also provided radial velocities for $\sim 7.2$ million bright stars \citep{gaia18}. It therefore offers an unprecedented opportunity to study the population of high-velocity stars in the Galaxy.

Based on their ejection mechanism and/or space velocities, high-velocity stars are usually divided in two different categories: runaway stars (RSs) and hypervelocity stars (HVSs), though the distinction is not always clear. The former likely originated in the Galactic disc and acquired high velocities either in supernova explosions in binary systems \citep{port00} or in close dynamical encounters involving stars and binaries \citep{gvar09,gvar11}. The latter, on the other hand, have such extreme velocities, often exceeding the local Galactic escape speed, that an interaction with the supermassive black hole (SMBH) in the Galactic Centre (GC) is required. The leading scenario, the classical Hills mechanism \citep{hills88}, involves the tidal disruption of a stellar binary by the SMBH which results in the capture of one star on a wide eccentric orbit and the ejection of the companion with very large velocity \citep{brm06,sari10}. Alternative mechanisms for the origin of HVSs include encounters with a massive black hole binary in the Galactic Centre \citep{yut03,fraglei18}, inspiralling intermediate-mass black holes \citep[IMBHs;][]{baum2006,sesa2006,sesan07}, encounters in a nearby galaxy \citep{GPZ2007,she2008,bou2017,erk19} and tidal interactions of stars clusters with a single or binary SMBHs \citep{cap15,fra16,fck17}. Unbound stars originating from the Galactic disc have also been discovered, e.g. HD 271791 \citep{heber08}, and these are often named hyperrunaway stars (HRSs).

More than $20$ early-type HVSs have been confirmed by the spectroscopic Multiple Mirror Telescope (MMT) survey, with Galactocentric velocities up to $\sim 700\kms$ and distances between $\sim 50$ and $\sim 120\kpc$ from the GC \citep{brw06,brw12,brw14}. Recently, \citet{boub18} used \textit{Gaia} data to show that late-type HVS candidates \citep[see e.g.][]{silva11,zhong14,vickers15} are likely bound to the Milky Way, except for LAMOST J115209.12+120258.0 \citep{li15}, which moves on an unbound orbit not originating in the GC.

Determining the exact origin of HVSs is of extreme importance for studies of the Galactic mass distribution and dark halo. If originating in the GC and travelling on almost radial orbits to the halo, HVSs can be used to probe the shape of the Galactic potential \citep{gnedin2010,fl2017,rossi2017}. \citet{brown18} show that only the fastest HVSs (with radial velocities $\gtrsim 450\kms$) have orbits originating in the GC, while the other unbound stars in the MMT sample have ambiguous origin. \citet{march18} used \textit{Gaia} data to identify HVS-candidates and found $28$ objects out of $165$ with a significant ($\gtrsim 50\%$) probability of being unbound, of which $\sim 2-5$ come from the GC and $\sim 9$ from the Galactic disc, and the remaining are likely of extragalactic origin. While HVSs coming from other galaxies and satellites (as the Large Magellanic Cloud; LMC) may contaminate the sample \citep{erk19}, \citet{keny18} showed that the Galactic disc and the LMC potential may have a role in deflecting HVSs from a nearly radial orbit. Yet, the origin of most of the known HVSs remains unknown \citep{brown18}.

A possible origin for the Galactic HVSs that cannot be traced back to the GC may be a Hills-type process in star clusters hosting an IMBH. Assuming that the observed $M_{SMBH}-\sigma$ relation (with $\sigma$ the local stellar velocity dispersion) holds also for the range of IMBH masses ($10^2\msun \lesssim M_{IMBH} \lesssim 10^5 \msun$), star clusters are the place where IMBHs should reside \citep*{fragk18,frlgk18}. 

The recent observation of a tidal disruption event consistent with an IMBH in an off-centre star cluster \citep{lin18} represents a significant milestone
in the hunt for the elusive IMBH population. In these clusters, binaries are disrupted by the tidal field of the IMBH and are ejected from the cluster with high velocities. 

In this paper, we study the ejection of HVSs from encounters between binary stars and an IMBH, as first proposed by \citet{pfa05} and \citet{GPZ2007}, in the core of star clusters by means of high-precision scattering experiments. Using a similar approach, \citet{sesan12} suggested that IMBH in star clusters may eject a population of millisecond pulsars to the Galactic halo. We study the imprint of the binary mass, binary semi-major axis and IMBH mass on the spatial and velocity distributions of the stars ejected from the cluster. Moreover, for the first time, we consider the effect of the cluster orbit through the Galaxy on the observable properties of the high-velocity stars, which can contribute both to the Galactic RS and HVS populations. We show that HVSs generated by a Hills-like mechanism in a star cluster hosting an IMBH would not travel on radial orbits consistent with a Galactic Centre origin, rather they would point back to their parent cluster, thus revealing the presence of an IMBH. We also show that high-velocity stars ejected by IMBHs have distinctive distributions in velocity, Galactocentric distance and Galactic latitude, which can be used to distinguish them from high-velocity stars from other channels.

The paper is organised as follows. In Section \ref{sect:hills}, we describe the classical Hills model for binary breakups by an IMBH. In Section \ref{sect:numsetup}, we describe the methods used in our numerical experiments. In Section \ref{sect:distributions}, we present the velocity distributions of HVSs after the tidal breakup of a binary by an IMBH, while, in Section \ref{sect:clusters}, we discuss the impact of the host cluster orbit on the HVS spatial distribution. In Section \ref{sect:comparisons}, we compare the properties of the high-velocity stars produced by the Galactic population of globular clusters to the high-velocity stars originated both in the GC and the Galactic disc. Finally, a discussion and conclusions are presented in Section \ref{sect:concl}.

\section{Binary interactions with an Intermediate-mass Black Hole}
\label{sect:hills}

\begin{figure} 
\centering
\includegraphics[scale=0.55]{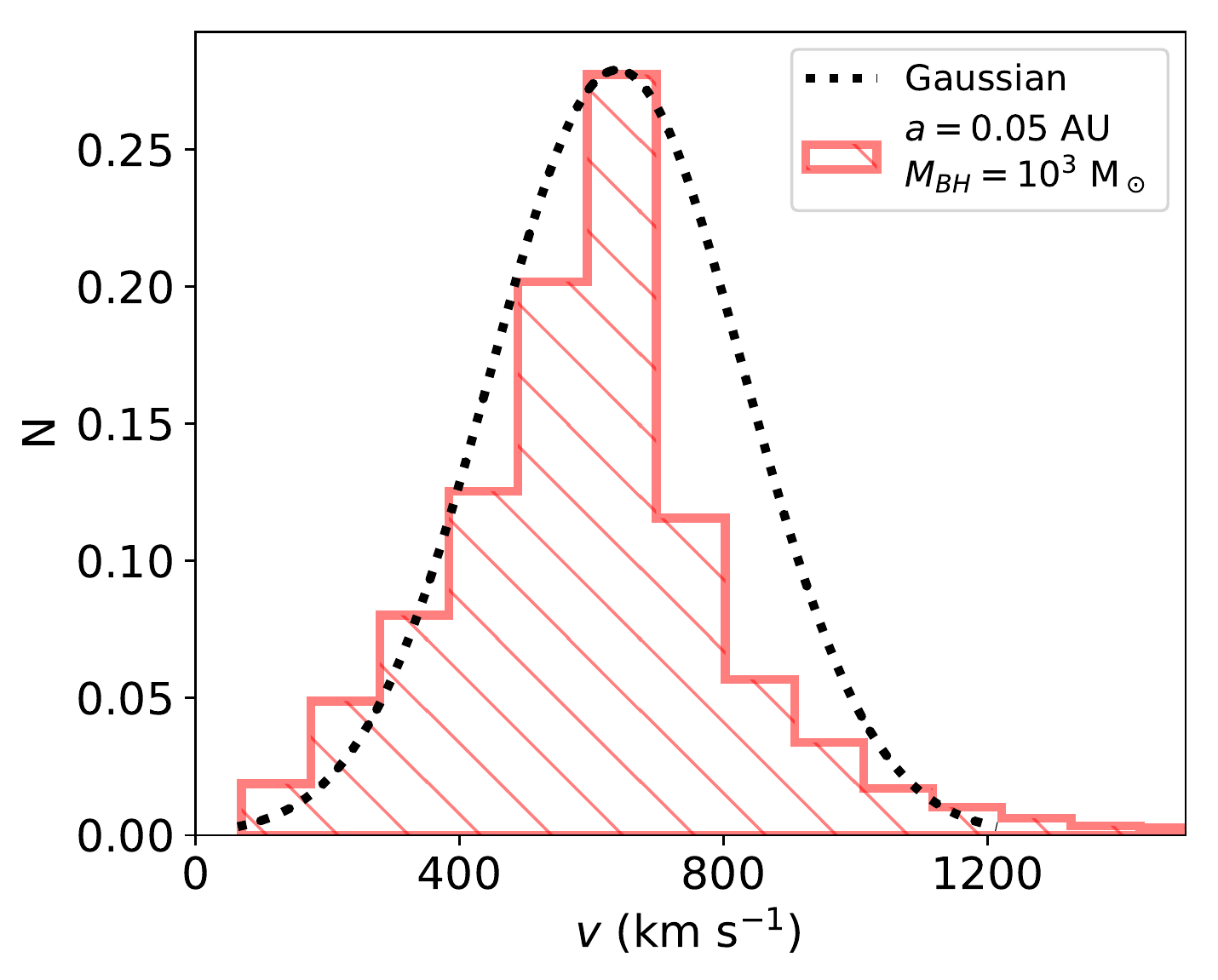}
\caption{Distribution of ejection velocities for stars in Model 1 (see Table\,\ref{tab1}) for $a=0.05\au$ (solid line). The dotted line represents a Gaussian distribution with mean $\mu = v_{ej}$, as given by Eq. \ref{eqn:vej}, and dispersion $\sigma=0.3\,v_{ej}$.}
\label{fig:gaussian}
\end{figure}

We consider a binary star of total mass $m=m_1+m_2$ and semi-major axis $a$ that undergoes a close interaction with an IMBH of mass $M$. In general, there are three possible outcomes for binary disruptions: (i) production of an ejected star and a star bound to the IMBH; (ii) production of $2$ stars bound to the IMBH; (iii) capture of the binary. Disruptions occur whenever the binary approaches the IMBH within the tidal radius
\begin{equation}
r_t\approx \left(\frac{M}{m}\right)^{1/3} a\ .
\label{eqn:rt}
\end{equation}
The typical velocity of the ejected star is \citep{brm06}
\begin{equation}
v_{ej}\approx 460 \left(\frac{a}{0.1\au}\right)^{-1/2} \left(\frac{m}{2\msun}\right)^{1/3} \left(\frac{M}{10^3\msun}\right)^{1/6} \kms\ .
\label{eqn:vej}
\end{equation}
For an unequal-mass binary, the ejection speeds of the primary and secondary are
\begin{equation}
v_1=v_{ej}\left(\frac{2m_2}{m}\right)^{1/2}\ ,
\label{eqn:vej1}
\end{equation}
\begin{equation}
v_2=v_{ej}\left(\frac{2m_1}{m}\right)^{1/2}\ ,
\label{eqn:vej2}
\end{equation}
respectively. The ejection speeds in the previous equations represent theoretical speeds at infinite distance from the IMBH and are averages over the initial phases and orientations of the binary. Figure \ref{fig:gaussian} illustrates the distribution of ejection speeds for an equal-mass binary ($m_1=m_2=1\msun$) with initial semi-major axis $a=0.05\au$ after encountering an IMBH of mass $M=10^3\msun$. The distribution is peaked at $\sim 700 \kms$, with tails extending to $\sim 1300 \kms$. The dashed line represents a Gaussian distribution with mean $v_{ej}$ as predicted by Eq. \ref{eqn:vej} and dispersion $0.3v_{ej}$, which provides a reasonable characterization of the numerical results, as already found in \citet{brm06}. The broadening of the distribution around the Hills peak is due to averaging over the random orientations of the binary orbit and orbital plane.

\section{Numerical setup}
\label{sect:numsetup}

\begin{table}
\caption{Models: name, mass of the IMBH ($M$), mass of the first star ($m_1$), mass of the second star ($m_2$), initial binary semi-major axis ($a$).}
\centering
\begin{tabular}{|l|c|c|c|c|}
\hline
Name & $M$ ($\msun$) & $m_1$ ($\msun$) & $m_2$ ($\msun$) & $a$ (AU)\\
\hline
Model 1		& $10^3$		& $1$	 	& $m_1$		& $0.05$-$1.00$\\
Model 2		& $10^2$-$10^5$	& $1$	 	& $m_1$		& $0.5$\\
Model 3		& $10^3$ 		& $1$-$8$   & $m_1$		& $0.5$\\
Model 4		& $10^4$ 		& $8$	 	& $1$-$8$	& $0.5$\\
\hline
\end{tabular}
\label{tab1}
\end{table}

We perform scattering experiments in which each binary starts from a distance $D=10^3 a$ from the IMBH. We generate the maximum initial distance for which the pericentre $\rmin$ of the binary is $\rmin\lesssim r_t$ \citep{fgu18,fgi18}. We then randomly sample initial distances up to such maximum according to a probability $f(b)\propto b$ in the pericentre distance, as appropriate when gravitational focusing is important \citep{hills88,brm06}.

The initial conditions for the numerical experiments have been set as follows  (see also Table\,\ref{tab1}):
\begin{itemize}
\item The mass of the IMBH is $M=(10^2,10^3,10^4,10^5)\msun$.
\item Stellar masses are set to $m_*=(1,2,4,8)\msun$.
\item Stellar radii are computed from \citep{dem91}
\begin{equation}
R_*=
\begin{cases}
1.06\ (m_*/\msun)^{0.945}\rsun& \text{$ m_*< 1.66\msun$},\\
1.33\ (m_*/\msun)^{0.555}\rsun& \text{$ m_*> 1.66\msun$}.
\end{cases}
\end{equation}
If the distance between stars becomes smaller than the sum of the stellar radii, the stars are considered merged and removed from the simulation.
\item The semi-major axis of the binary is $a=0.05$-$1.0\au$.
\item The initial eccentricity of the binary is $e=0$.
\item The initial phase $\chi$ of the binary, which determines the initial position of the stars in the orbit, is randomly generated.
\item The angle $\theta$, which determines the relative inclination of the binary's centre-of-mass orbit with respect to the IMBH and the binary orbital plane is randomly generated.
\item The initial distance of the binary from the IMBH is $D=10^3 a$.
\end{itemize}
The minimum and maximum semi-major axis are set by requiring that the stars do not collide in their first orbit nor are they unbound by scatering with cluster stars in the core, respectively \citep{frasar18}. Initial circular orbits are not a serious limitation, given that the results depend mainly on the initial binary energy content\citep{brm06}. 

We integrate the system for a total time $T=D/v$, where $v$ is the initial velocity of the centre of mass of the binary, which we set to $v=10\kms$ as appropriate for the core of a cluster hosting an IMBH. This choice of $T$ allows us to follow all encounters 
to completion, i.e. until a formal classification of the outcome based on energy considerations can be performed.

We integrated the system of the differential equations of motion of the 3-bodies
\begin{equation}
{\ddot{\textbf{r}}}_i=-G\sum\limits_{j\ne i}\frac{m_j(\textbf{r}_i-\textbf{r}_j)}{\left|\textbf{r}_i-\textbf{r}_j\right|^3}\ ,
\end{equation}
with $i=1$,$2$,$3$, using the \textsc{archain} code \citep{mik06,mik08}, a fully regularised code able to model the evolution of systems of arbitrary mass ratios with extreme accuracy, even over long periods of time. 

\section{Velocity distributions of ejected stars}
\label{sect:distributions}

\begin{figure} 
\centering
\includegraphics[scale=0.55]{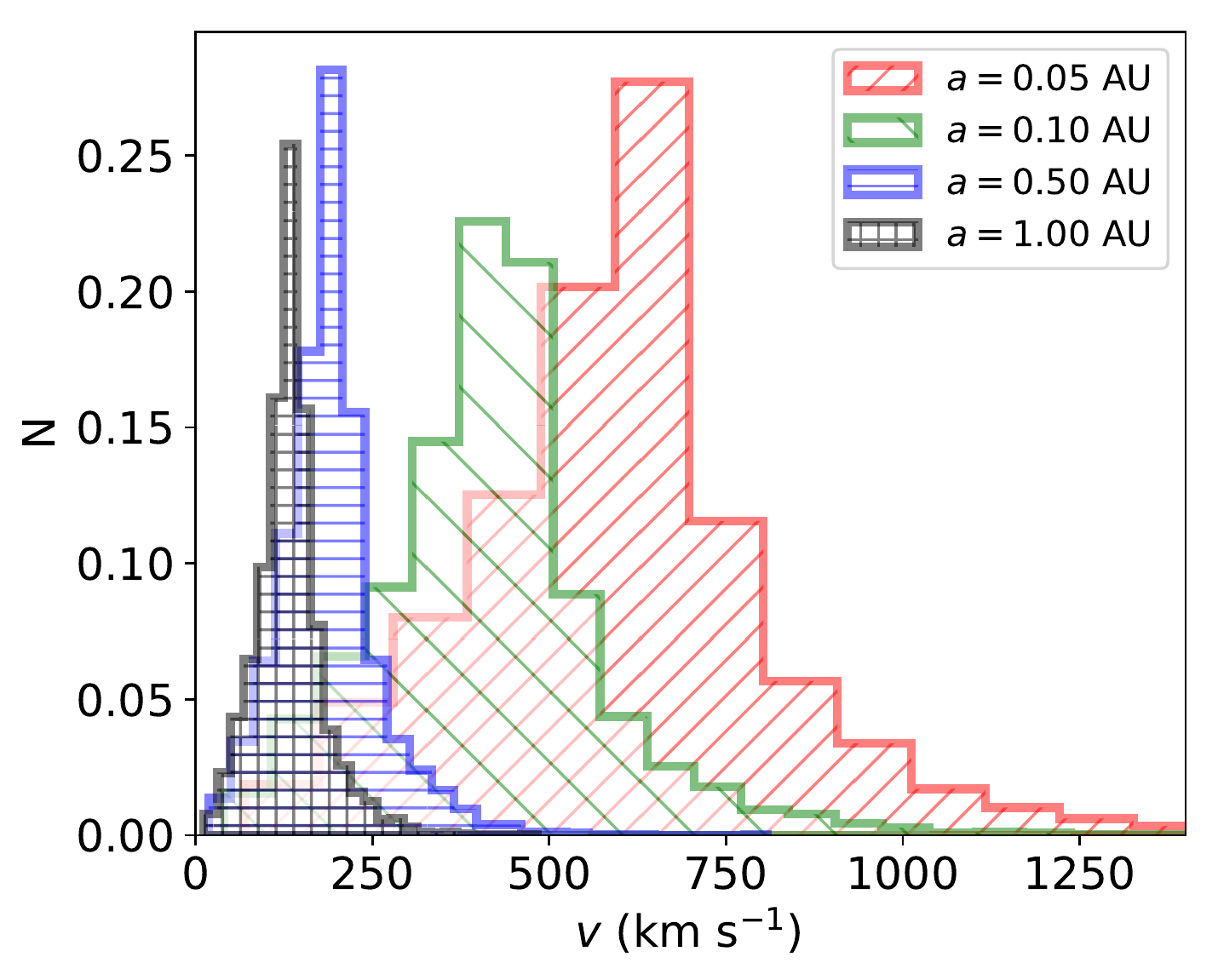}
\caption{Distribution of ejection velocities for stars in Model 1 for $M=10^3\msun$, $m_1=m_2=1\msun$ and different initial binary semi-major axes.}
\label{fig:model1}
\end{figure}

\begin{figure} 
\centering
\includegraphics[scale=0.55]{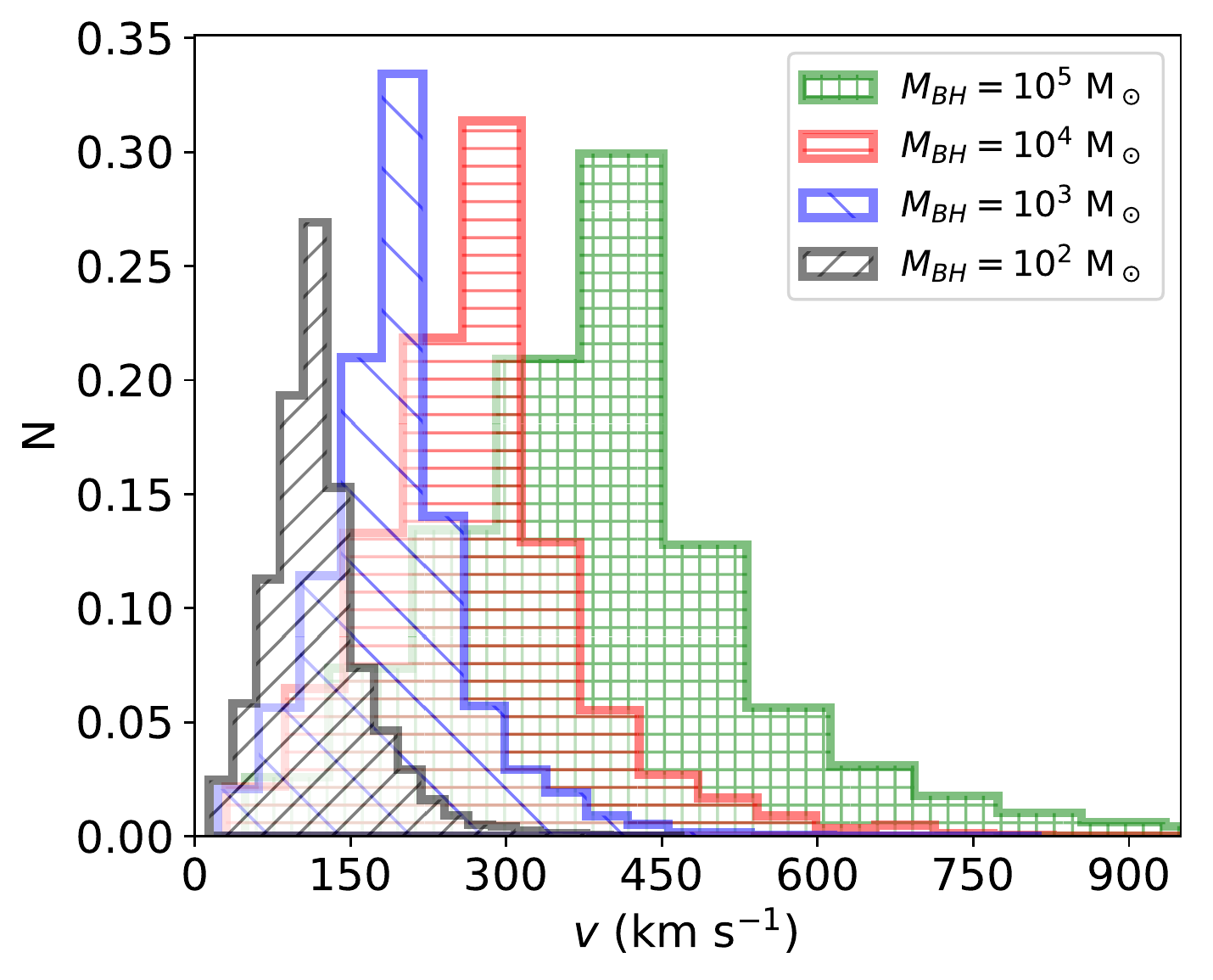}
\caption{Distribution of ejection velocities for stars in Model 2 for equal-mass binaries with $m_1=m_2=1\msun$ and initial semi-major axis $a=0.5\au$, for different values of the IMBH mass.}
\label{fig:model2}
\end{figure}

\begin{figure*} 
\centering
\begin{minipage}{20.5cm}
\includegraphics[scale=0.55]{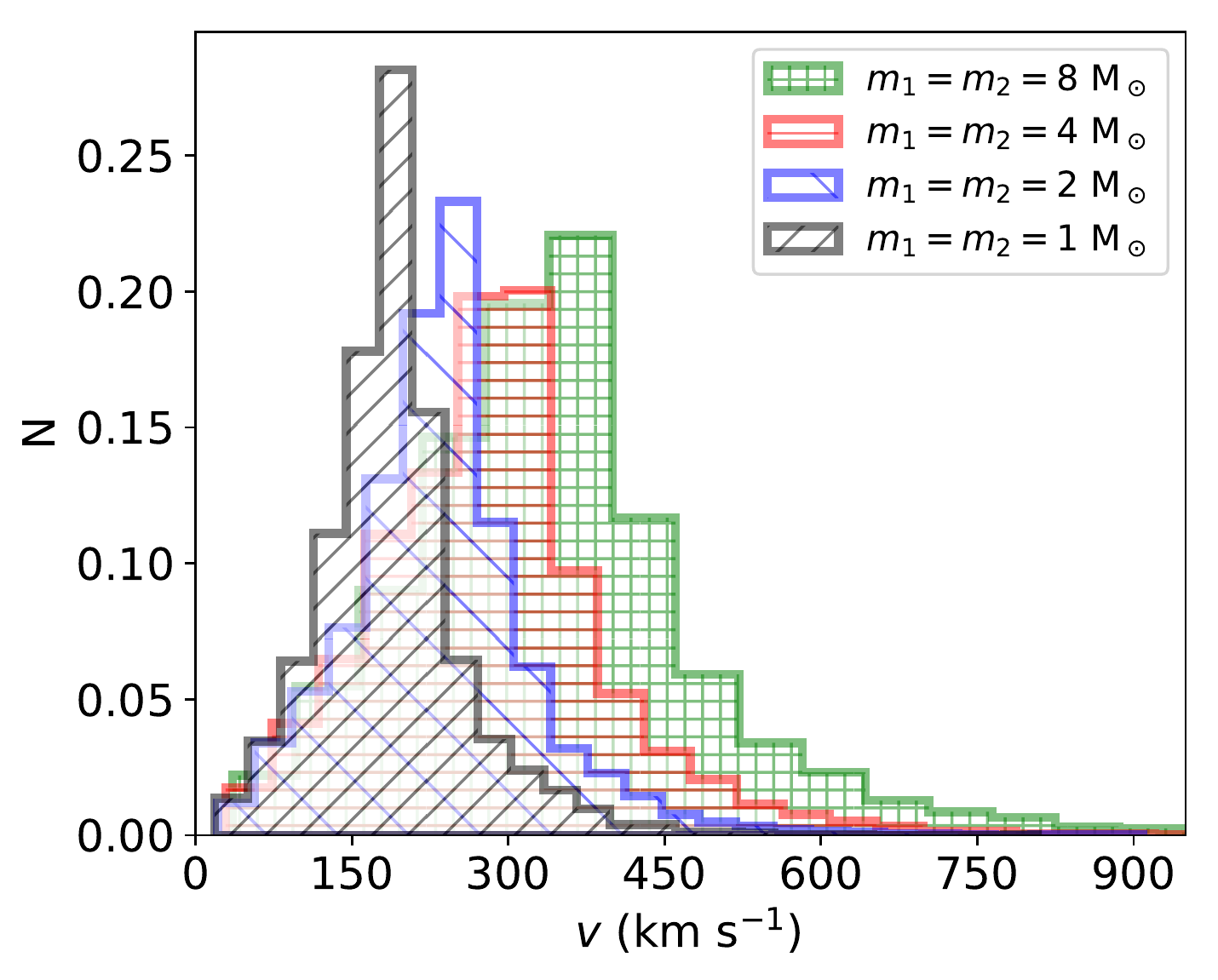}
\hspace{1cm}
\includegraphics[scale=0.55]{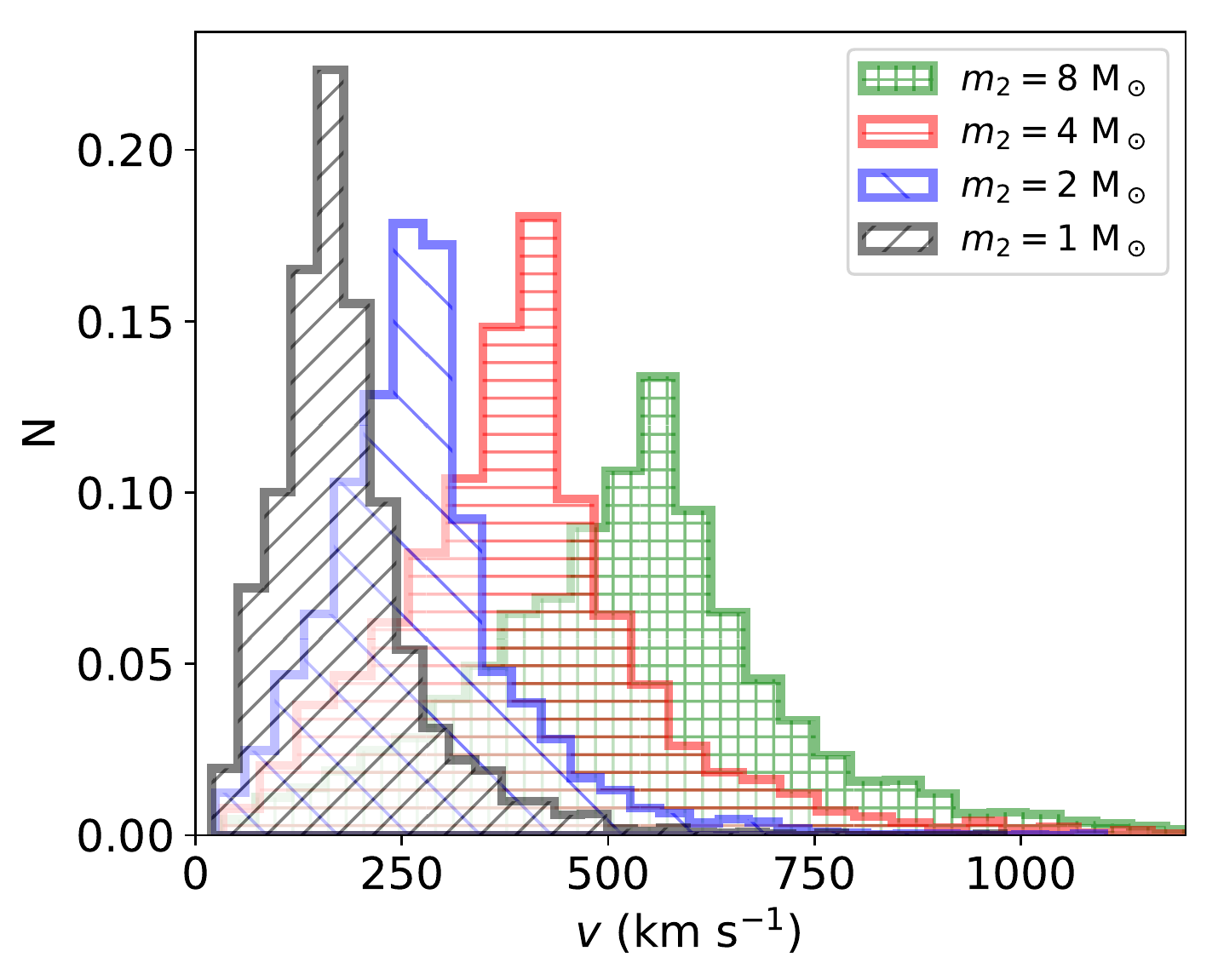}
\end{minipage}
\caption{Left: Velocity distribution of the ejected stars in Model 3 for $M=10^3\msun$, initial binary semi-major axis $a=0.5\au$ and different stellar masses $m_1=m_2$. Right: Velocity distribution of the ejected stars in Model 4 for $M=10^4\msun$, initial binary semi-major axis $a=0.5\au$, $m_1=8\msun$ and different masses of the companion star $m_2$.}
\label{fig:model34}
\end{figure*}

We performed $5000$ simulations of IMBH-stellar binary close encounters for each combination of the parameters given in Table\,\ref{tab1}, for a total of $80000$ experiments. 

In Model 1, we study the fate of binaries as a function of the initial semi-major axis $0.05\au\le a\le 1.00\au$, while fixing $M=10^3\msun$ and $m_1=m_2=1\msun$.  We show the distribution of the ejection velocities in Fig. \ref{fig:model1}. The distributions are peaked at $\sim 160\kms$, $\sim 200\kms$, $\sim 500\kms$ and $\sim 700\kms$ for $a=0.05\au$, $0.10\au$, $0.50\au$ and $1.00\au$, respectively. The peak values are consistent with the Hills peak $v_{ej}$ of Eq. \ref{eqn:vej}, and the distributions are well described by a Gaussian with mean $v_{ej}$ and dispersion $\sigma=0.3\,v_{ej}$. As predicted in Eq. \ref{eqn:vej}, the smaller the initial semi-major axis, the largest the peak velocity. Moreover, larger initial binary semi-major axes imply a more pronounced broadening around the peak value.

In Model 2, we examine the velocity distributions of the ejected stars as a function of the IMBH mass, taken to be $M=(10^2,10^3,10^4,10^5)\msun$, and consider equal-mass binaries with $m_1=m_2=1\msun$ and initial semi-major axis $a=0.5\au$. The distributions in Fig. \ref{fig:model2} are peaked at $\sim 120\kms$, $\sim 200\kms$, $\sim 280\kms$ and $\sim 440\kms$ for $M=10^2\msun$, $M=10^3\msun$, $10^4\msun$ and $10^5\msun$, respectively. Similarly to Model 1, the peak values are consistent with the theoretical Hills peak $v_{ej}$ from Eq. \ref{eqn:vej}, showing that the mass of the IMBH has much less impact on the peak value of the velocity distribution compared to the initial binary semi-major axis and mass, being $v_{ej}\propto M^{1/6}$. We also found that a larger IMBH mass implies a larger broadening of the velocity distributions around $v_{ej}$.

\begin{figure*} 
\centering
\begin{minipage}{20.5cm}
\includegraphics[scale=0.55]{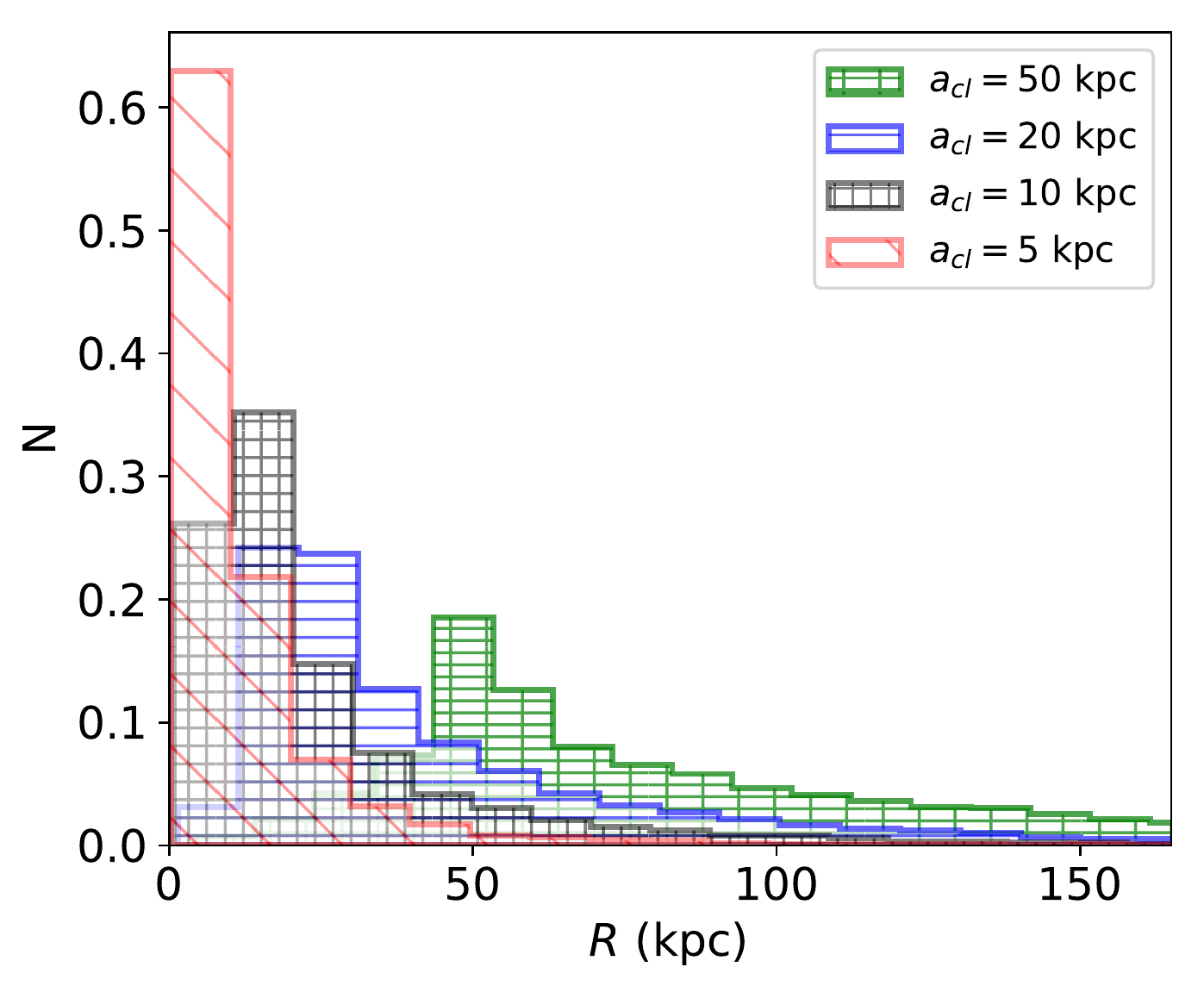}
\hspace{1cm}
\includegraphics[scale=0.55]{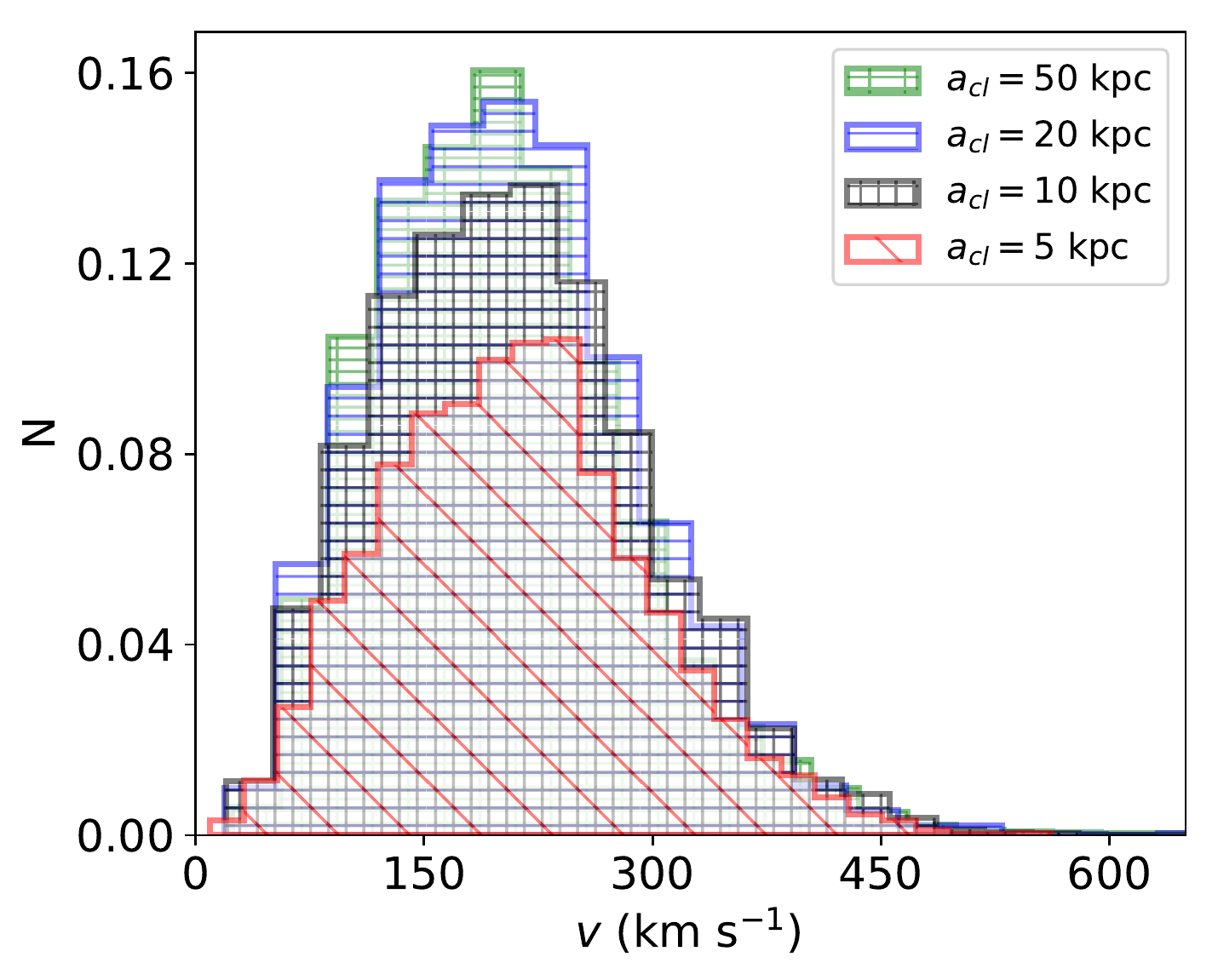}
\end{minipage}
\begin{minipage}{20.5cm}
\includegraphics[scale=0.55]{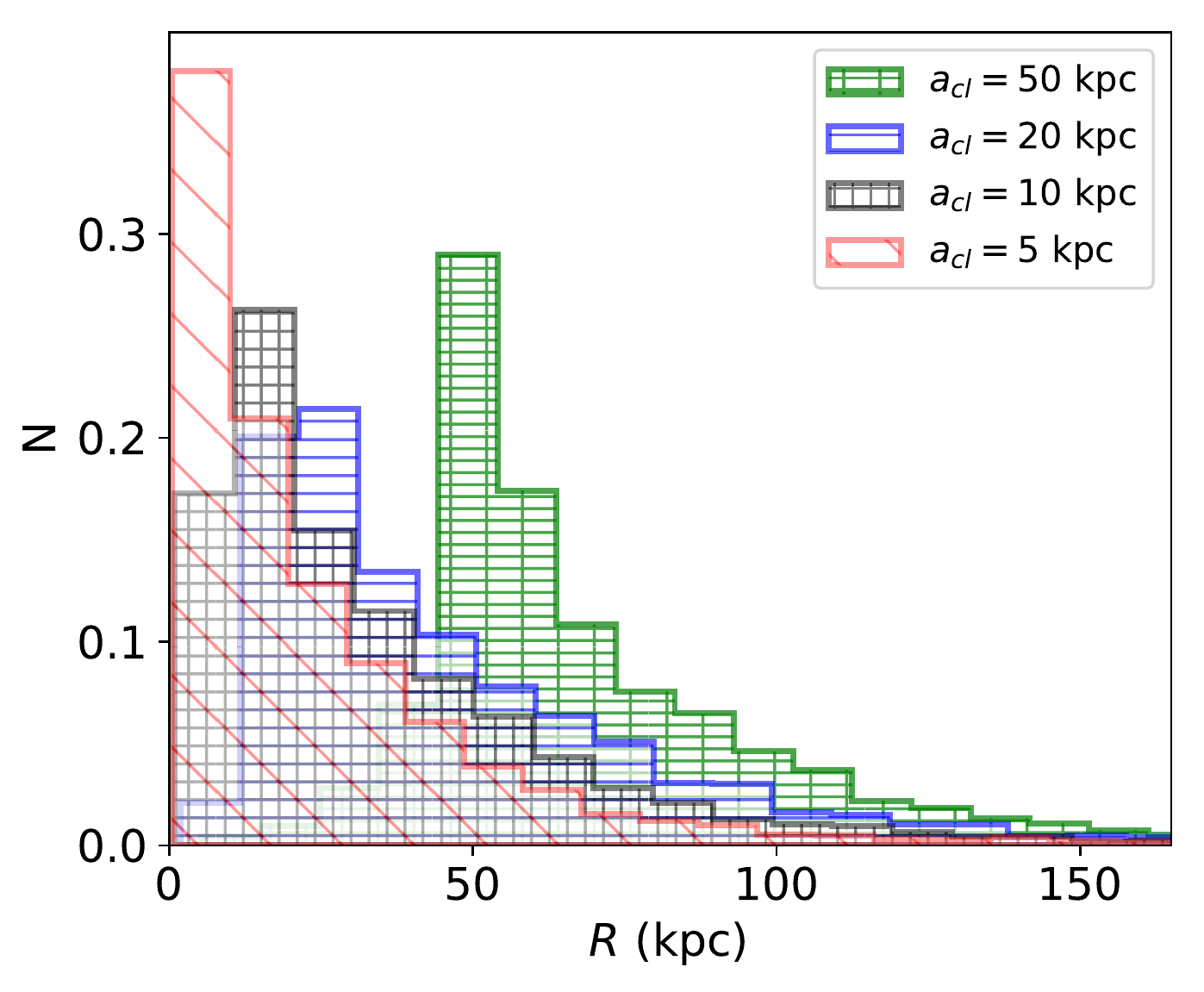}
\hspace{1cm}
\includegraphics[scale=0.55]{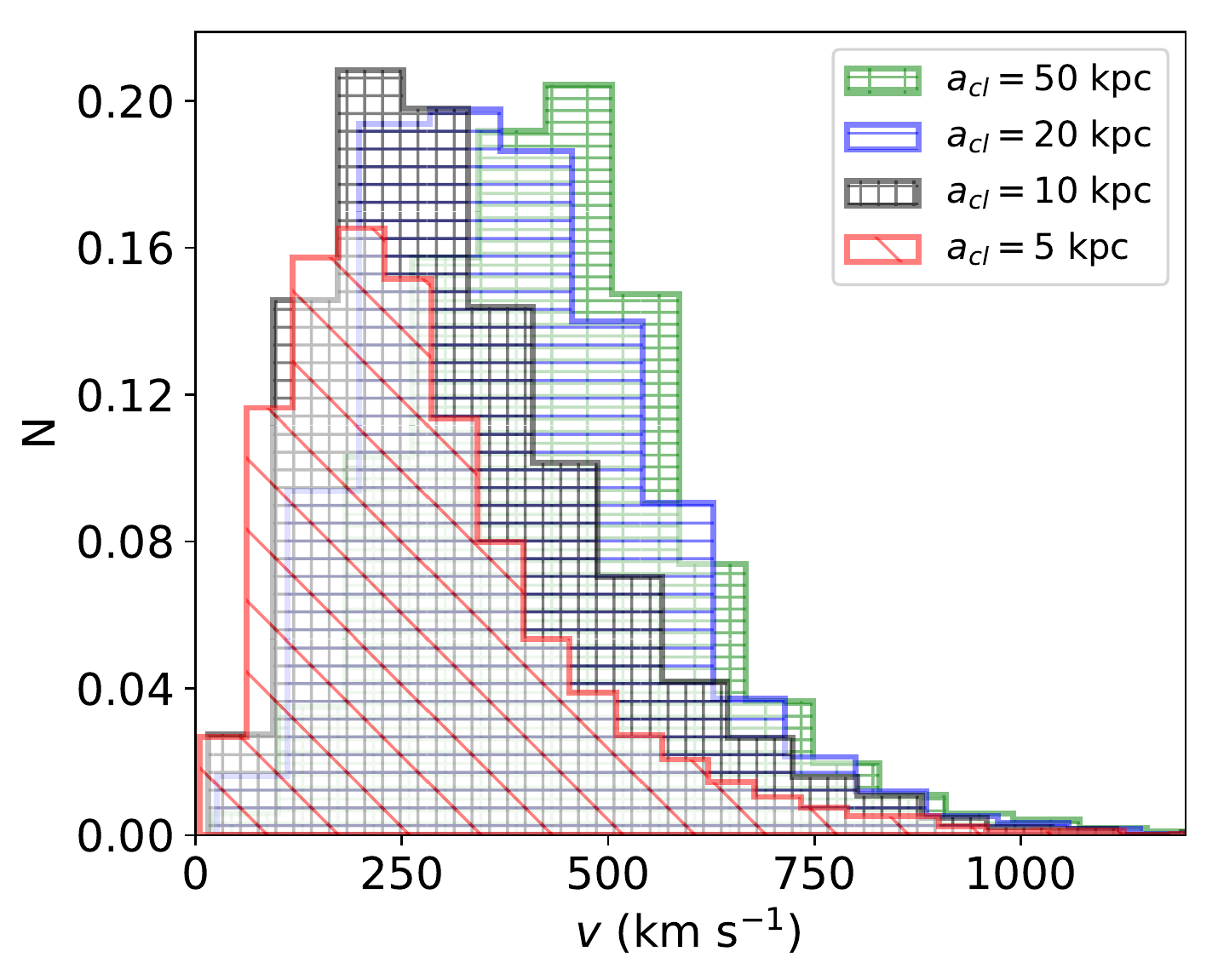}
\end{minipage}
\caption{Galactocentric distance (left) and velocity (right) distribution of the stars ejected from a cluster hosting an IMBH of mass $M=10^4\msun$ after a binary-IMBH interaction of equal-mass binaries of mass $m_1=m_2=1\msun$ (top) and $m_1=m_2=4\msun$ (bottom), with $a=0.5\au$. The host cluster orbits the GC on a circular orbit of semi-major axis $a_{cl}$ in the Galactic disc.}
\label{fig:cluster}
\end{figure*}

In Model 3 and Model 4, we consider the role of the mass of the stars in the binaries in shaping the velocity distribution of the ejected stars. We fix the mass of the IMBH to $M=10^3\msun$ (Model 3) and $M=10^4\msun$ (Model 4), and the initial binary semi-major axis to $a=0.5\au$. In Model 3, we consider equal-mass binaries in the mass range $1\le m_1=m_2\le 8$, while in Model 4 we fix $m_1=8\msun$ and consider $1\le m_2\le 8$ for the mass of the secondary. The left panel in Fig.\,\ref{fig:model34} shows the velocity distribution of the ejected stars in Model 3 as function of the mass of the binary components. The distributions are peaked at $\sim 200\kms$, $\sim 270\kms$, $\sim 340\kms$ and $\sim 400\kms$ for $m_1=m_2=1\msun$, $m_1=m_2=2\msun$, $m_1=m_2=4\msun$ and $m_1=m_2=8\msun$, respectively, showing a dependence $v_{ej}\propto m^{1/3}$ in agreement with Eq.\,\ref{eqn:vej}. The right panel in Fig.\,\ref{fig:model34} shows the velocity distribution of the ejected stars in Model 4 as function of the mass of the secondary star $m_2$ in the binary. The distributions are peaked at $\sim 190\kms$, $\sim 290\kms$, $\sim 420\kms$ and $\sim 600\kms$ for $m_2=1\msun$, $m_2=2\msun$, $m_2=4\msun$ and $m_2=8\msun$, respectively. The peaks are in agreement with Eqs. \ref{eqn:vej}-\ref{eqn:vej1}. 
As previously, the larger the binary mass the broader the velocity distribution.

\section{Ejecting hypervelocity stars from star clusters}
\label{sect:clusters}

The ejection of stars from encounters between an IMBH and a stellar binary
is expected to take place in the core of star clusters, where an IMBH with mass in the range $10^2\msun \lesssim M \lesssim 10^5 \msun$ could reside \citep*{fragk18,frlgk18}. When a star is ejected from a cluster, its ejection velocity needs to be combined with the cluster orbital velocity at the time of ejection.

To compute the Galactic distribution of stars ejected from a cluster hosting an IMBH, we use the results of the scattering experiments and combine the ejection velocity of the stars with the cluster velocity at the moment of ejection. We describe the cluster orbit by means of its semi-major axis $a_{cl}$. The cluster eccentricity $e_{cl}$ and the relative inclination of its orbital plane with respect to the Galactic disc $\eta_{cl}$ do not affect the final velocities of the stars, but only their spatial distribution, being the velocity of the cluster along its orbit at maximum of the same order as the ejection velocity. Assuming a constant ejection rate, a large orbital eccentricity for the cluster would imply that the majority of the stars are ejected near the cluster apocentre. Moreover, stars ejected from the cluster tend to reside in the cluster orbital plane, whose orientation is determined by $\eta_{cl}$. In our calculations, we have assumed $e_{cl}=0$ and $\eta_{cl}=0^\circ$.

We describe the Galactic potential with a 4-component model $\Phi=\Phi_{SMBH}+\Phi_{b}+\Phi_{d}+\Phi_{h}$ \citep{ken14,fl2017}, where:
\begin{itemize}
\item $\Phi_{BH}$ is the contribution of the central SMBH,
\begin{equation}
\Phi_{SMBH}(r)=-\frac{GM_{SMBH}}{r},
\end{equation}
with $M_{SMBH}=4\times 10^6$ M$_{\odot}$;
\item $\Phi_{b}$ is the contribution of the spherical bulge,
\begin{equation}
\Phi_{b}(r)=-\frac{GM_{b}}{r+a_b},
\end{equation}
with mass $M_{b}=3.76\times 10^9$ M$_{\odot}$ and core radius $a_b=0.10$ kpc;
\item $\Phi_{d}$ accounts for the axisymmetric disc,
\begin{equation}
\Phi_d(R,z)=-\frac{GM_{d}}{\sqrt(R^2+(b+\sqrt{c^2+z^2})^2)},
\end{equation}
with mass $M_{d}=5.36\times 10^{10}\msun$, length scale $b=2.75$ kpc and scale height $c=0.30$ kpc;
\item $\Phi_h$ is the contribution of the dark matter halo
\begin{equation}
\Phi_h(r)=-\frac{GM_{h}\ln(1+r/r_s)}{r}.
\end{equation}
with mass $M_{h}=10^{12}\msun$ and length scale $r_s=20$ kpc.
\end{itemize}
The parameters are chosen so that the Galactic circular velocity at the distance of the Sun ($8.15$ kpc) is $235\kms$.

To generate mock catalogs of stars, we assume that the central IMBH ejects stars at a constant rate along the cluster orbit. We follow the formalism of \citet{pfa05} to compute the rate of ejections from a cluster due to encounters with an IMBH. We distinguish between full and empty loss-cone, which is the region in phase-space where binaries interact with the IMBH and can be disrupted. In the full loss-cone regime, binary stars may be scattered in and out of the loss-cone on their way from apoapse to periapse, while, in the empty loss-cone regime, any star deflected into the loss-cone is disrupted within a dynamical time. \citet{pfa05} found for the full loss-cone regime
\begin{eqnarray}
\mathcal{R} & \approx & f_b \left(\frac{a}{0.1 \au}\right) \left(\frac{n}{10^5 \pc^3}\right) \left(\frac{M}{10^3 \msun}\right)^{4/3}\times\nonumber\\
& \times & \left(\frac{\sigma}{10\kms}\right)^{-1} \mathrm{Myr}^{-1}\,,
\end{eqnarray}
while, in case of empty loss cone
\begin{equation}
\label{eq:imbhrate}
\mathcal{R}\approx f_b \left(\frac{n}{10^5 \pc^3}\right)^{2} \left(\frac{M}{10^3 \msun}\right)^{3} \left(\frac{\sigma}{10\kms}\right)^{-9} \mathrm{Myr}^{-1}\, .
\end{equation}
In the previous equations, $f_b$ is the binary fraction, $\sigma$ is the velocity dispersion and $n$ the density of the host cluster. Both rates  predict approximately one binary disruption per Myr, thus we fix the ejection rate to $1$ Myr$^{-1}$ in our calculations.

To compute the observational properties of the ejected stars, we follow the procedure outlined in \citet{ken14}. For each ejected star we randomly choose an ejection time $t_{ej}$ and an observation time $t_{obs}$ between zero and its main-sequence lifetime $t_{ms}$. This assumes that the ejected star is observed before it evolves off the main-sequence. Whenever $t_{ej}<t_{obs}$, we assign the star an ejection velocity sampled from the results of our scattering experiments. We then combine the star ejection velocity with the cluster orbital velocity assuming a random ejection direction and assign the location of the cluster along its orbit as the star's position.
We then integrate the full 3D orbit of the star in the Galactic potential up to a maximum time $T=t_{obs}-t_{ej}$, tracking its position and velocity. If at any time the star reaches the virial radius (assumed to be $250$ kpc), the star is considered ejected from the Galaxy and removed from the calculation.

\begin{figure} 
\centering
\includegraphics[scale=0.55]{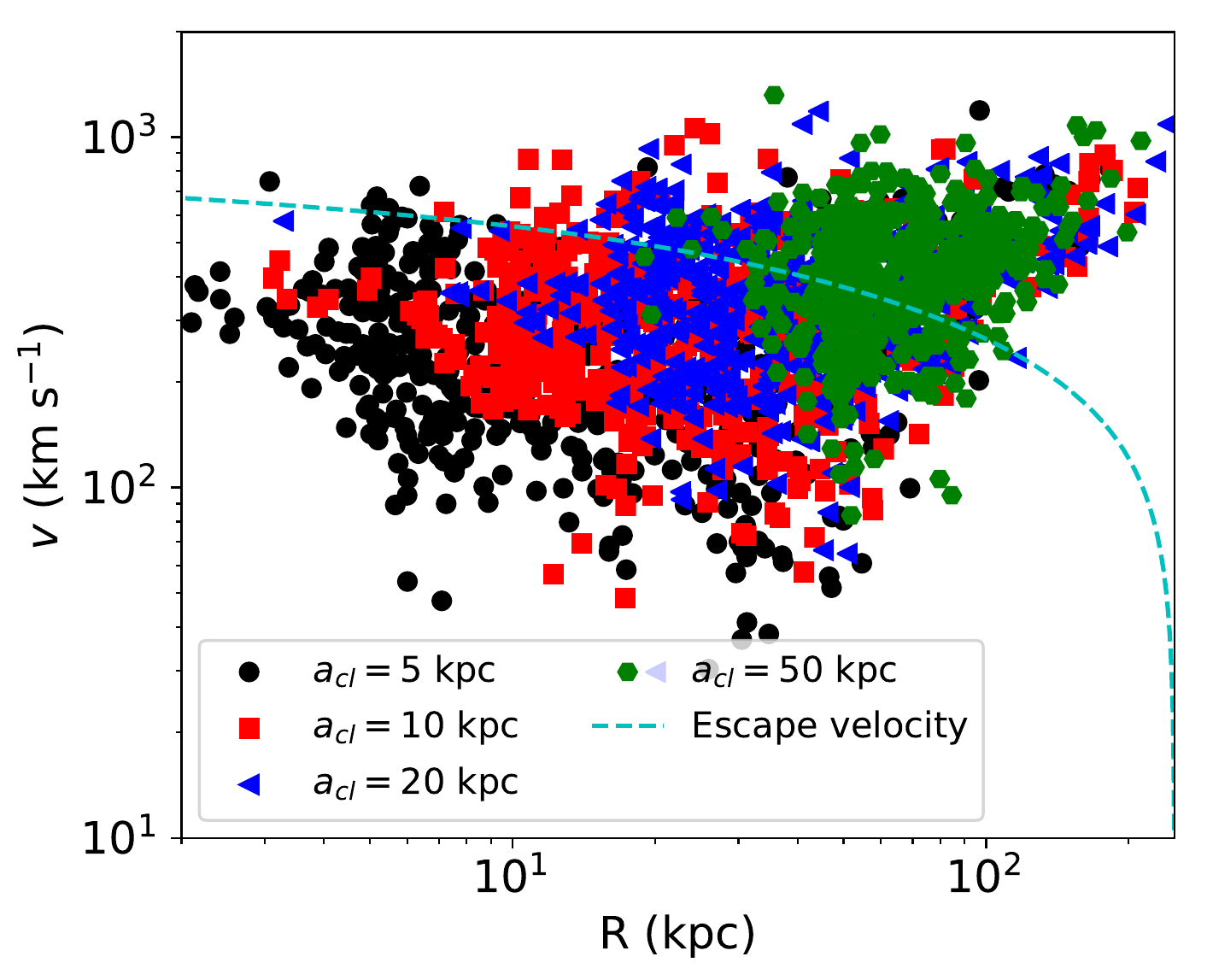}
\caption{Galactocentric velocity versus Galactocentric distance of $4\msun$ stars ($a=0.5\au$ and $M=10^4\msun$). Most of the unbound HVSs are found at $R\gtrsim 30$ kpc.  Clusters with larger $a_{cl}$ produce a larger number of unbound stars, being the local escape speed smaller.}
\label{fig:veldistm4}
\end{figure}

\begin{figure} 
\centering
\includegraphics[scale=0.55]{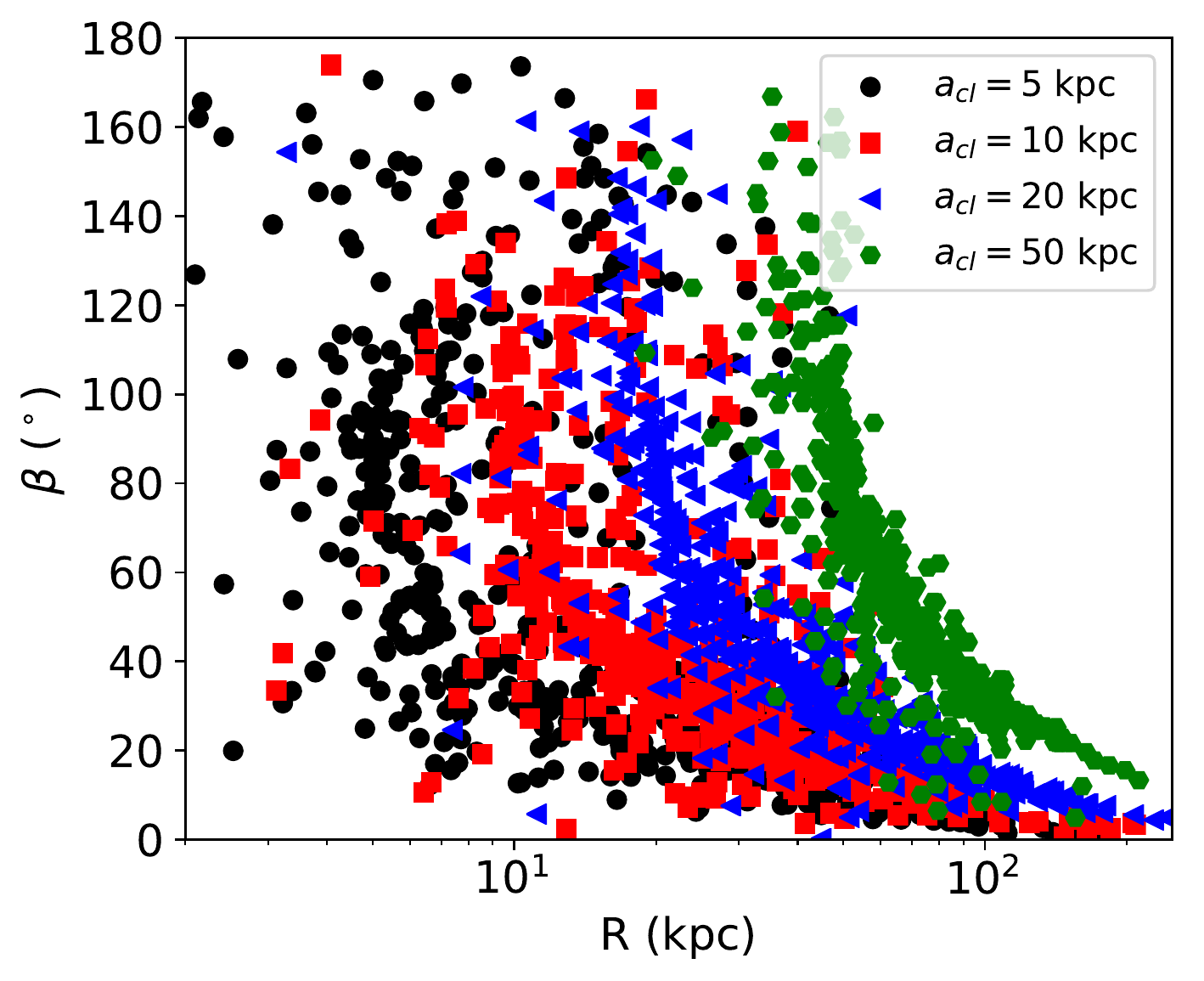}
\caption{Relative angle between the Galactocentric distance and velocity vectors (see Eq.\,\ref{eq:beta}) for $4\msun$ stars ($a=0.5\au$ and $M=10^4\msun$). Almost all stars have $\beta\gtrsim 0^\circ$, i.e. non-radial orbits with respect to the Galactic Centre.}
\label{fig:anglesm4}
\end{figure}

\begin{figure*} 
\centering
\begin{minipage}{20.5cm}
\includegraphics[scale=0.55]{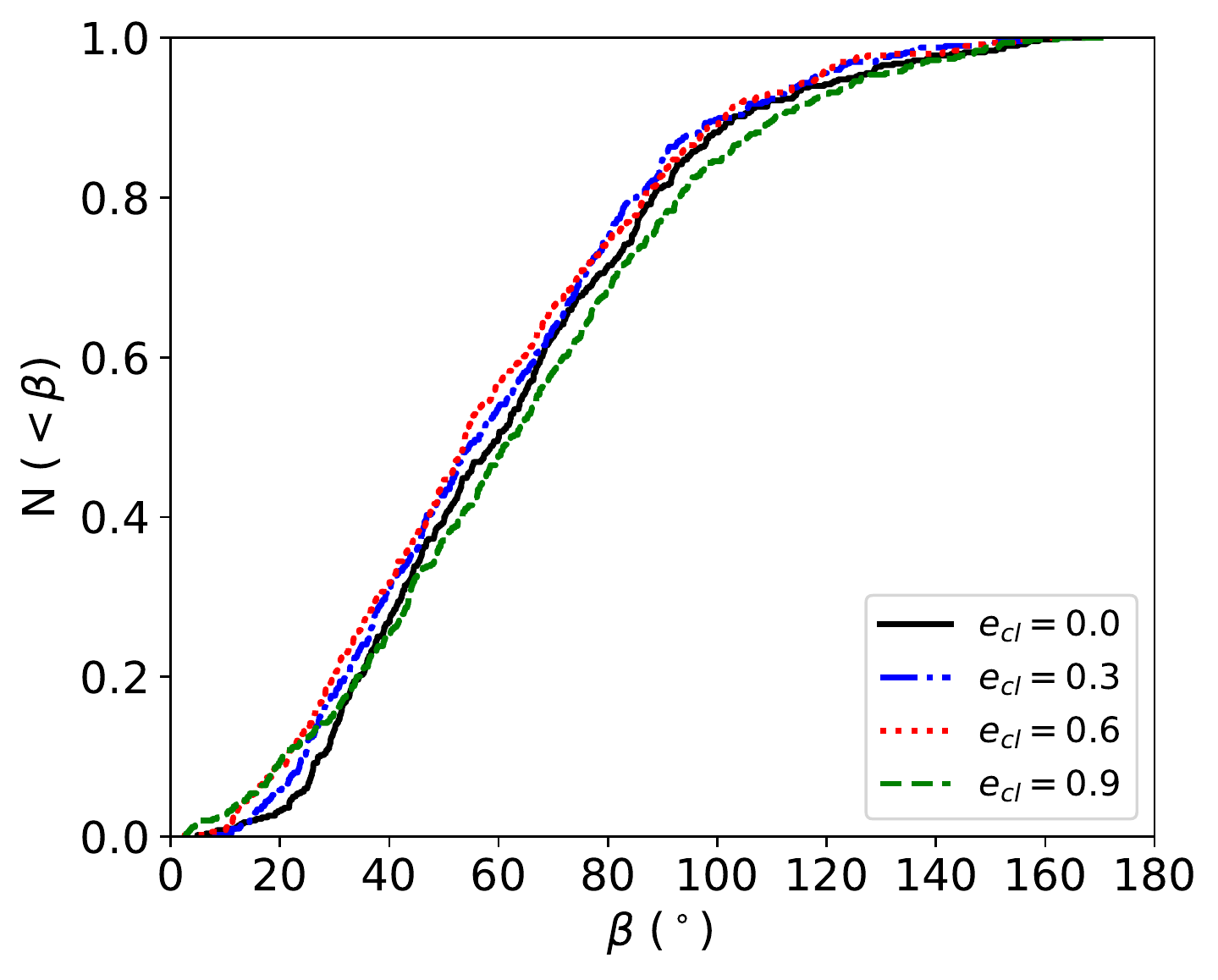}
\hspace{1cm}
\includegraphics[scale=0.55]{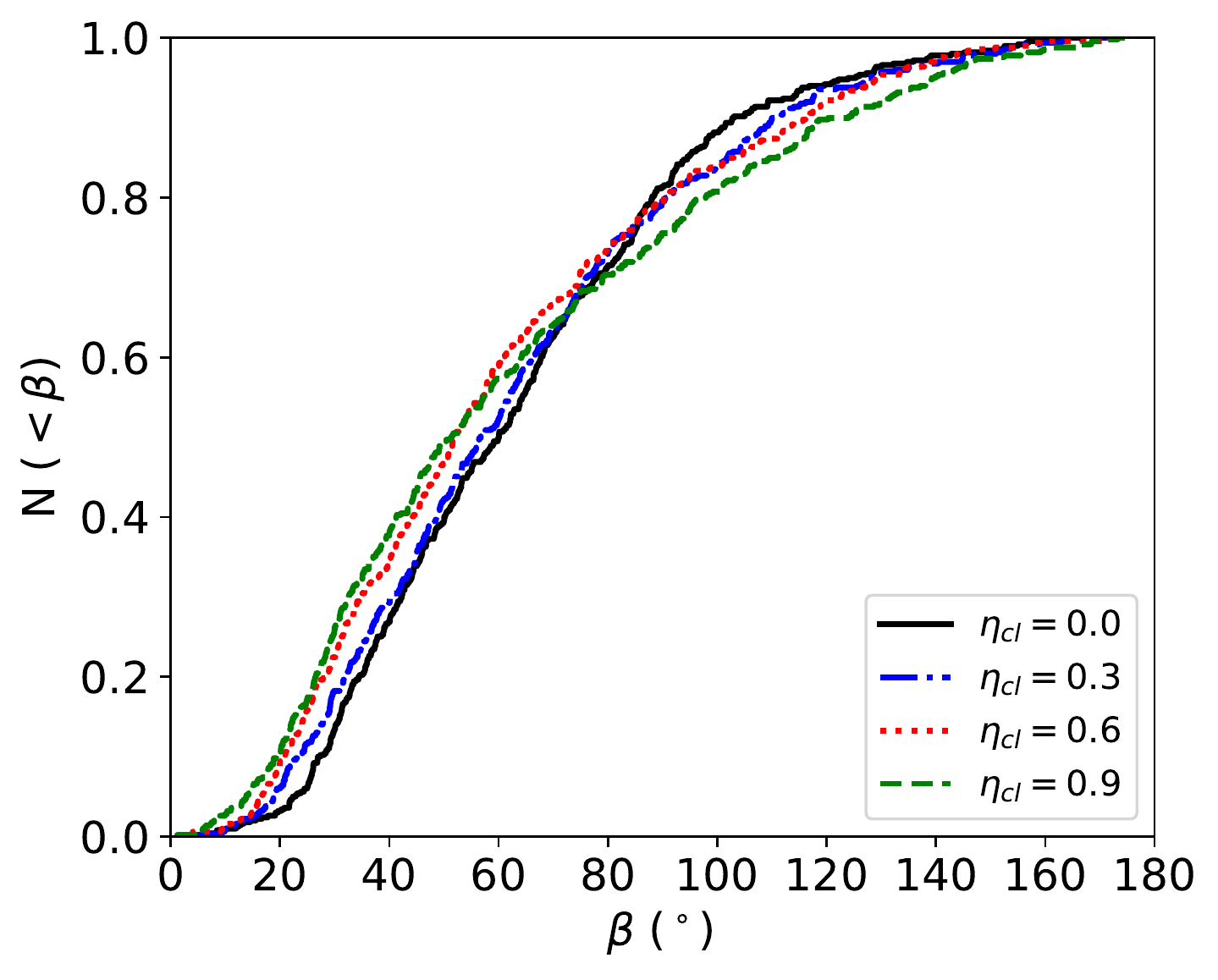}
\end{minipage}
\caption{Cumulative distributions of the relative angle $\beta$ between the Galactocentric distance and velocity vectors for $4\msun$ stars ($a=0.5\au$ and $M=10^4\msun$) and for $a_{cl}=50$ kpc, as a function of the cluster eccentricity (left) and inclination with respect to the Galactic disc (right).}
\label{fig:angles_ecceta}
\end{figure*}

\begin{figure*} 
\centering
\begin{minipage}{20.5cm}
\includegraphics[scale=0.55]{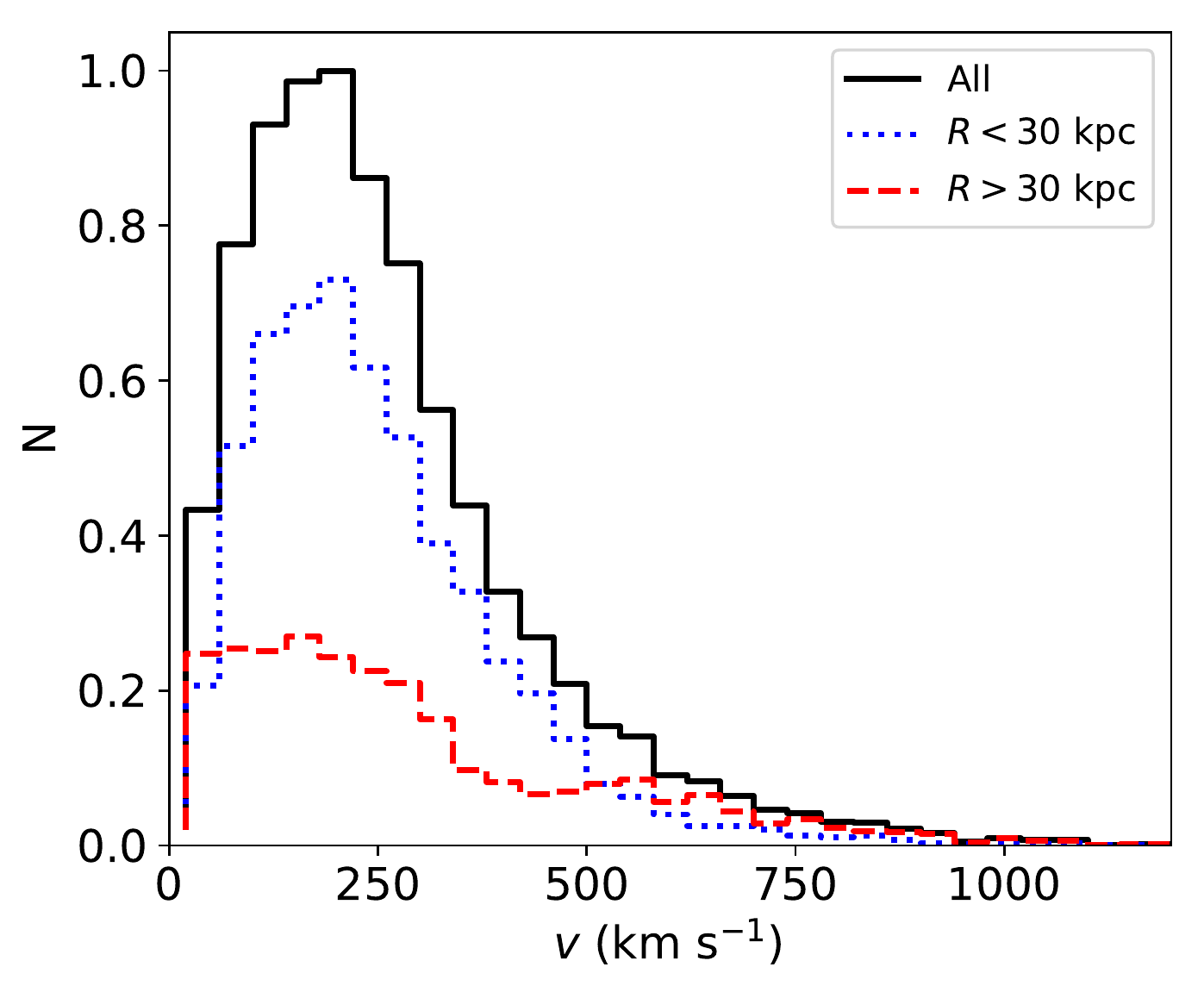}
\hspace{1cm}
\includegraphics[scale=0.55]{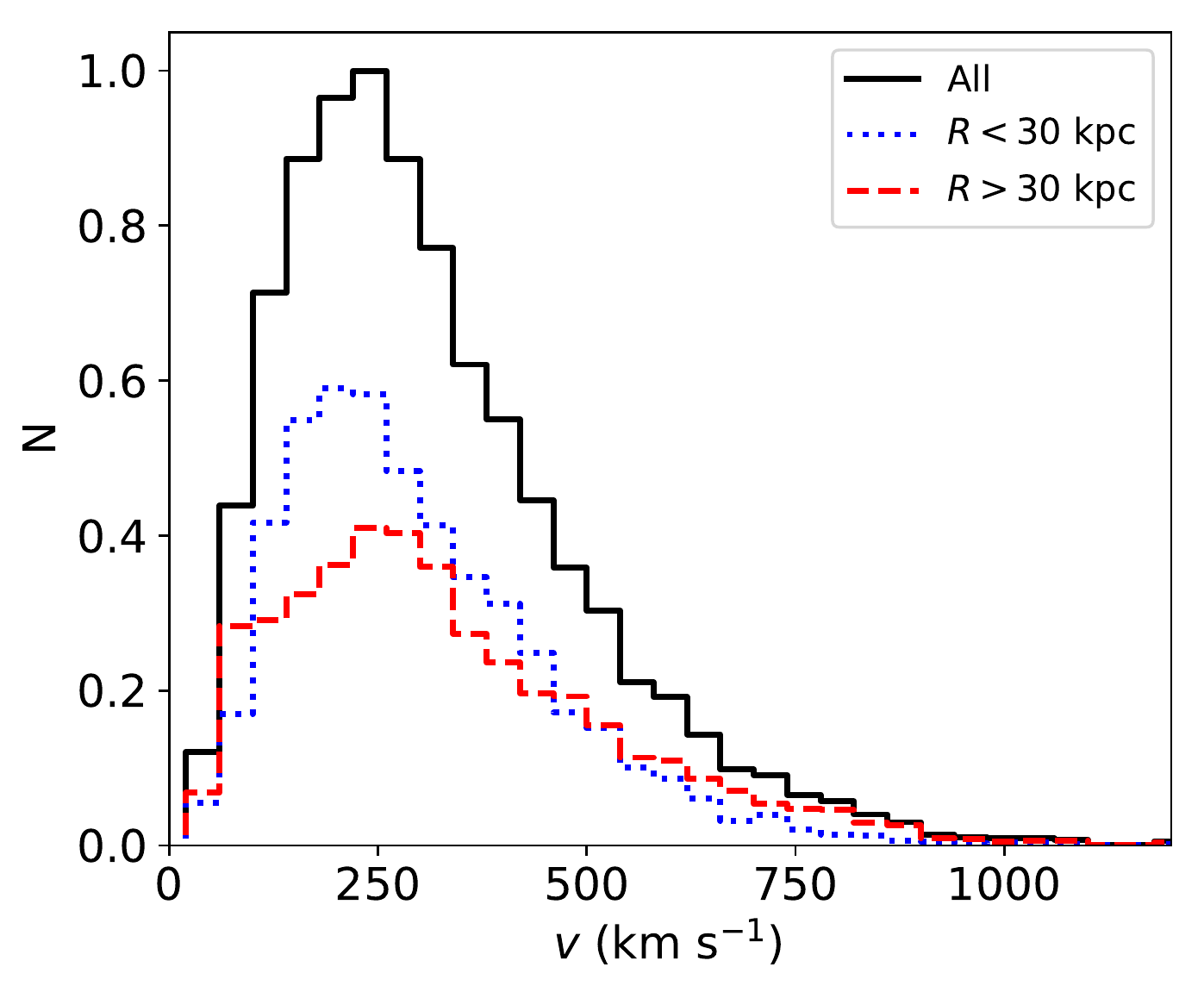}
\end{minipage}
\begin{minipage}{20.5cm}
\includegraphics[scale=0.55]{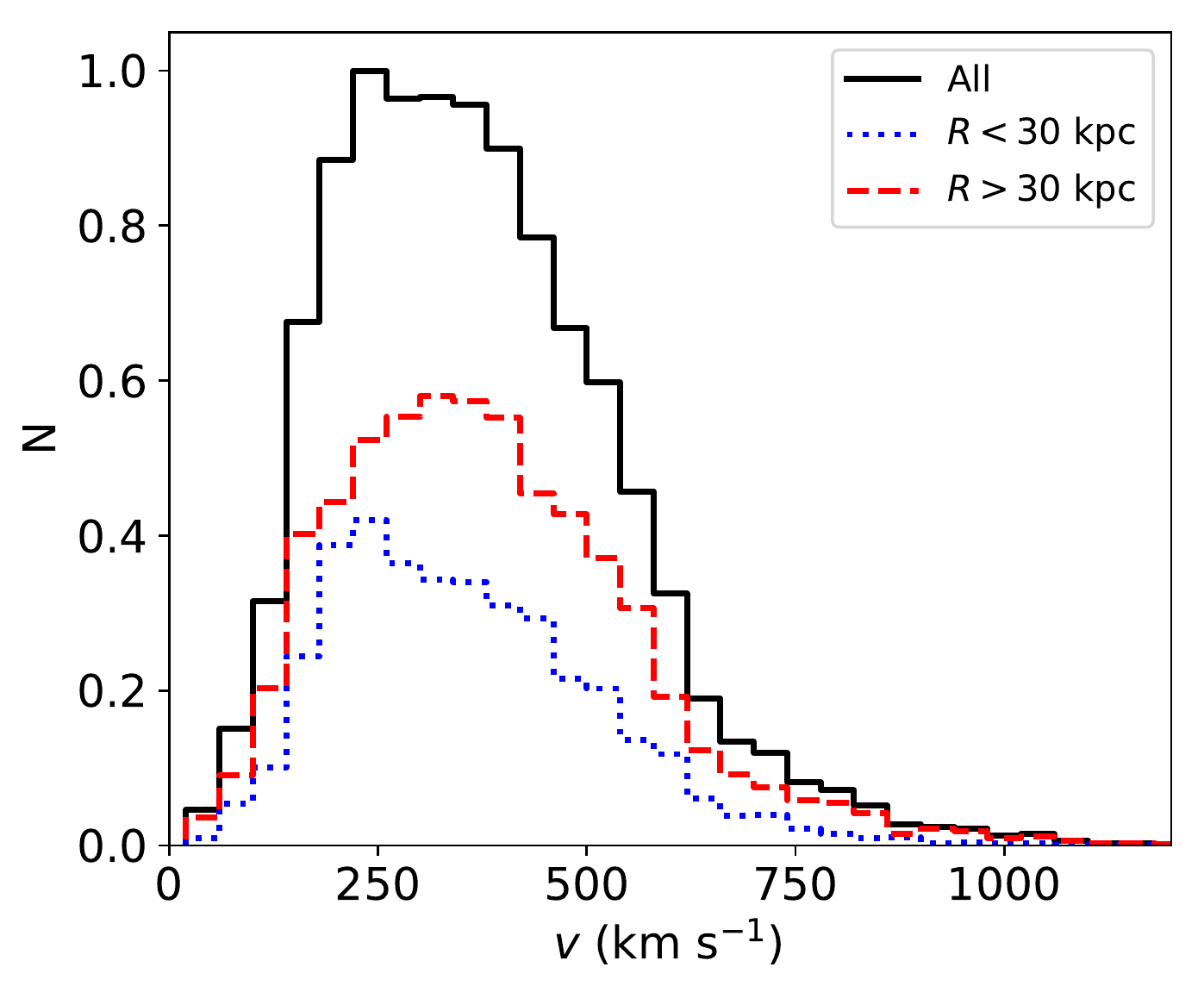}
\hspace{1cm}
\includegraphics[scale=0.55]{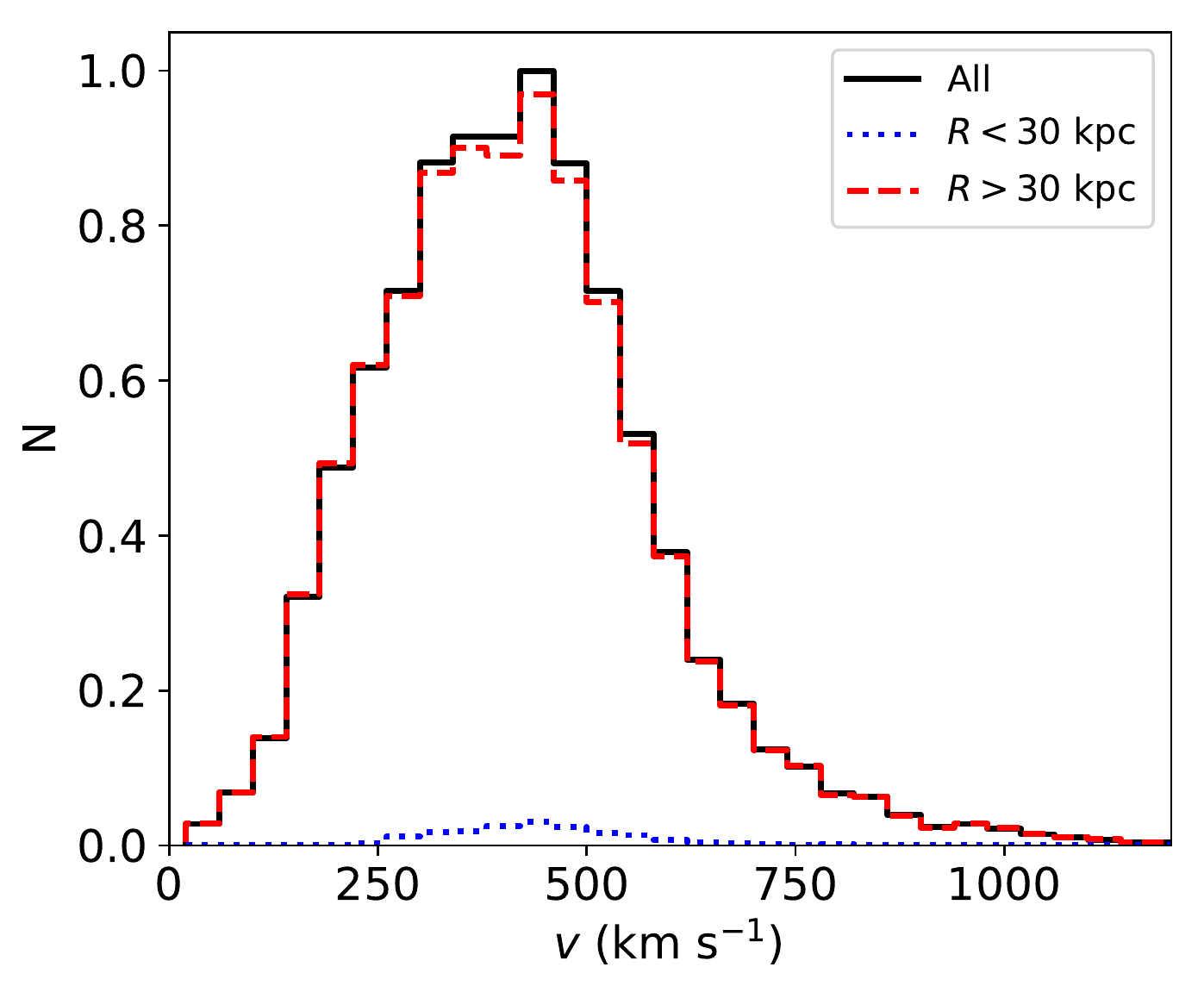}
\end{minipage}
\caption{Galactocentric velocity distribution of stars, originally in an equal-mass binary ($m_1=m_2=4\msun$, $a=0.5\au$), ejected from a cluster hosting an IMBH of mass $M=10^4\msun$ orbiting the GC on a circular orbit (in the Galactic disc) of semi-major axis $a_{cl}$=5 kpc (top-left), 10 kpc (top-right), 20 kpc (bottom-left) and 50 kpc (bottom-right). Blue-dotted lines represent stars within $30$ kpc from the GC, while red-dashed lines show stars outside this radius.}
\label{fig:detect}
\end{figure*}

Figure \ref{fig:cluster} shows the Galactocentric distance (left) and velocity (right) distribution of stars of mass $1\msun$ (top) and $4\msun$ (bottom), originally in an equal-mass binary with semi-major axis $a=0.5\au$, ejected from a cluster hosting an IMBH of mass $M=10^4\msun$ orbiting the GC on a circular orbit of semi-major axis $a_{cl}=(5,10,20,50)$ kpc in the Galactic disc. Regardless of the mass of the stars, the Galactocentric spatial distribution is peaked near the cluster semi-major axis, with tails extending to the Galactic virial radius. The mass of the stars affects, instead, the velocity distribution. For $1\msun$ stars, the velocity distribution is peaked at $\sim 200\kms$ independently of the cluster orbit, while the velocity distribution is peaked at $\sim 200\kms$ and $\sim 500\kms$ for $4\msun$ stars in the cases $a_{cl}=5$ kpc and $a_{cl}=50$ kpc, respectively. The dependence on stellar mass may be interpreted both in terms of the ejection velocities and in terms of the main-sequence lifetimes, computed as \citep{fl2017}
\begin{equation}
t_{ms}=\left(\frac{m}{1 \msun}\right)^{-2.5} 10^{10}\yr\ .
\end{equation}
Typically, $4\msun$ stars have a main-sequence lifetime ($t_{ms}=3\times 10^{8}$ yr) $\sim 30$ times smaller than $1\msun$ stars ($t_{ms}=10^{10}$ yr). Typically $1\msun$ stars are ejected with velocities $\sim 1.6$ times smaller than the $4\msun$ stars. In our configuration ($m_1=m_2=1\msun$ and $a=0.5\au$), the typical ejection velocity is $\sim 230\kms$, which has to be combined with the cluster orbital velocity. Moreover, $1\msun$ stars ejected with velocities exceeding the local escape speed ($\sim 650\kms$ and $\sim 350\kms$ at $5$ kpc and $50$ kpc, respectively) have sufficient time to travel out of the Galaxy within their main-sequence lifetime. This result is independent of the cluster orbit. As a consequence, only $1\msun$ stars with velocities below the local escape speed can be observed in the Galaxy. On the other hand, $4\msun$ stars have ejection speeds $\sim 370\kms$, but a smaller $t_{ms}=3\times 10^{8}$ yr. These stars can only be observed soon after ejection from the cluster both because they may evolve off the main-sequence and because they may escape the Galaxy. As a consequence, $4\msun$ stars have a much wider spectrum of velocities.

In Fig. \ref{fig:veldistm4}, we show the Galactocentric distances and velocities of $4\msun$ stars for different cluster semi-major axes. Most of the unbound HVSs have $R\gtrsim 30$ kpc, and clusters with larger $a_{cl}$ produce a larger number of unbound stars, being the local escape speed smaller. To help illustrate the deflection form a purely radial Galactocentric orbit of the high-velocity stars produced as a consequence of binary disruptions by an IMBH in a star cluster, we compute the angle
\begin{equation}
\label{eq:beta}
\beta=\arccos\left(\frac{\textbf{R}\cdot \textbf{v}}{Rv}\right)\ ,
\end{equation}
where $\textbf{R}$ and $\textbf{v}$ are the observable Galactocentric distance and velocity vectors, respectively. The case $\beta=0^\circ$ corresponds to purely radial orbits. Fig.\,\ref{fig:anglesm4} shows $\beta$ as a function of the Galactocentric distance for different cluster orbits. The vast majority of stars have non-radial orbits with respect to the GC. 

In Fig.\,\ref{fig:angles_ecceta}, we illustrate the effect of the cluster orbital eccentricity and inclination, in the case $a_{cl}=50$ kpc. As discussed above, the cluster eccentricity $e_{cl}$ and the relative inclination of its orbital plane with respect to the Galactic disc $\eta_{cl}$, do not affect the final Galactocentric velocity of the ejected stars, but do influence their spatial distribution. 

Figure \ref{fig:detect} shows the different contributions of stars within and outside $30$ kpc to the overall Galactocentric velocity distribution in the case of $4\msun$ stars ($a=0.5\au$ and $M=10^4\msun$). While for $a_{cl}=5$ kpc and $a_{cl}=10$ kpc, the majority of the stars is found within $30$ kpc, the stars outside this radius become the largest component for $a_{cl}=20$ kpc and $a_{cl}=50$ kpc. In the latter case, only $\sim 3\%$ of the stars is within $30$ kpc.

\section{High velocity stars ejected from the Milky Way Globular Clusters}
\label{sect:comparisons}

\begin{figure} 
\centering
\includegraphics[scale=0.6]{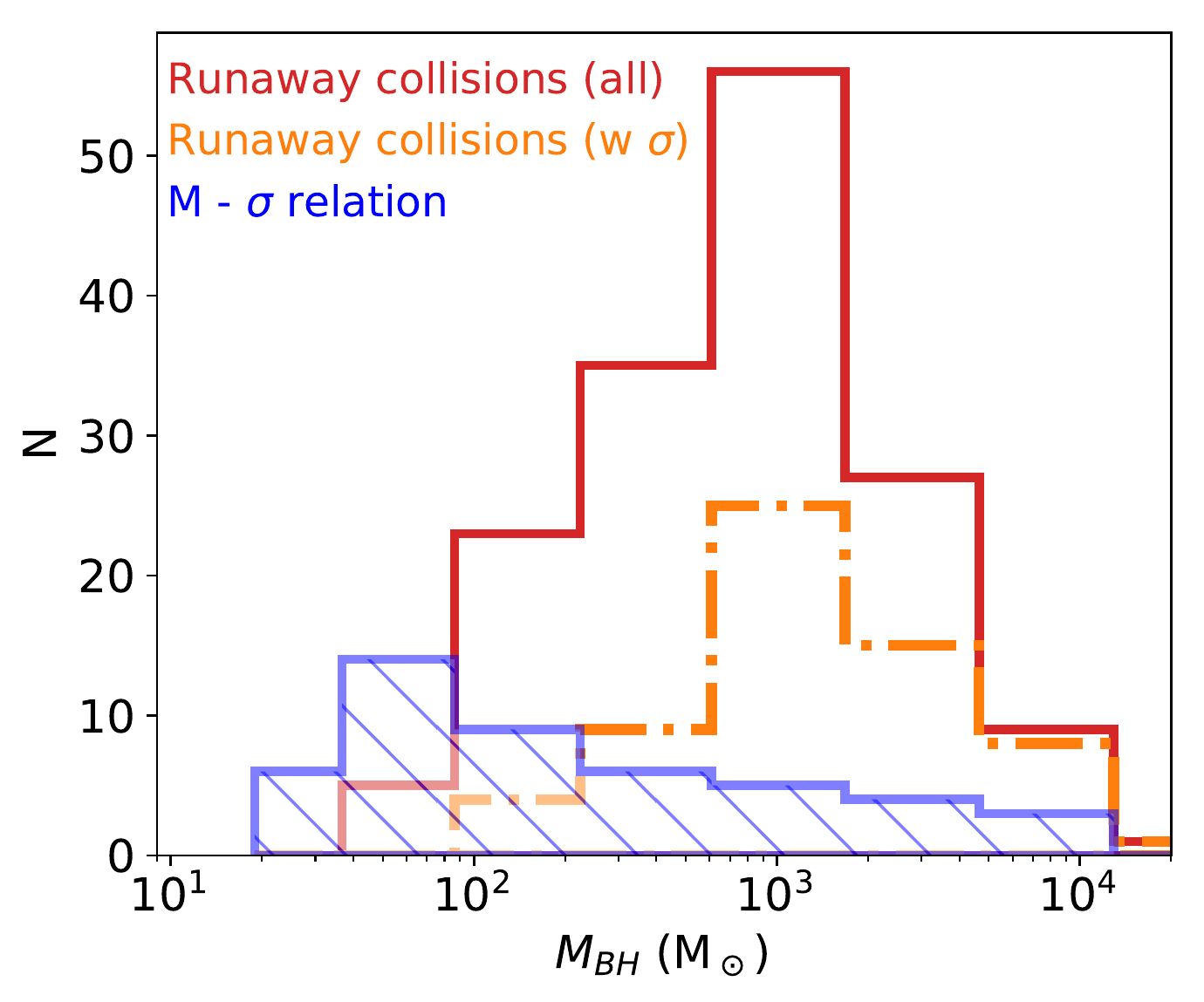}
\caption{Distribution of IMBH masses in the Milky Way's globular clusters adopting data from the \citep[][version 2010]{harr96} catalogue and considering both an extrapolation of the $M-\sigma$ relation to lower velocities (blue/shaded histogram) and a scaling from runaway collisions (red/empty histogram). Since not all clusters have a measured velocity dispersion, we also show the distribution of IMBH masses for the subsample of cluster with a measured value of $\sigma$.}
\label{fig:imbhmassmw}
\end{figure}

The Milky Way hosts about 150 globular clusters, and about 60 of them have reliable measurements of the 3D cluster velocity and the stellar velocity dispersion \citep[][version 2010]{harr96}. We can therefore model the ejection of high velocity stars from the Galactic population of globular clusters assuming they host an IMBH at the centre, and trace their orbits in the Galactic potential. Following \citet{sesan12}, we estimate the mass of the IMBH in each globular cluster in two different ways. In the first model, we assume that the IMBH is formed through runaway collisions of massive stars in the cluster core. The collisions build up a supermassive star which then collapses to a black hole \citep{pormc02}. The IMBH mass is found to scale with the cluster mass $M_{\rm GC}$ as
\begin{equation}
M_{BH}=m_{\rm seed}+4\times 10^{-3} f_c M_{\rm GC} \ln \Lambda\ ,
\label{eqn:scal}
\end{equation}
where $m_{\rm seed}\approx 50\msun$ is the mass of the heavy star that initiates the runaway process, $f_c\approx 0.2$ is the runaway coefficient and $\ln \Lambda \sim 10$ is the Coulomb logarithm \citep{pormc02}. In the second model, we set the IMBH mass from an extrapolation of the $M-\sigma$ relation to lower velocities \citep{trem02} 
\begin{equation}
M_{BH}=2 \left(\frac{\sigma}{70\kms}\right)^4 10^6\msun\ ,
\label{eqn:msig}
\end{equation}
where $\sigma$ is the cluster stellar velocity dispersion. 
The resulting distributions of IMBH masses are shown in Fig.~\ref{fig:imbhmassmw}. We notice that the runaway collisions model produces significantly larger IMBH masses than the $M-\sigma$ model, with a peak velocity around $1000\msun$. The distribution of masses from the $M-\sigma$ model is peaked below $100\msun$ but is broader, with a long tail extending to $\sim 10^4\msun$. Since not all clusters have a measured value of $\sigma$, we also consider the subsample of clusters with a measured velocity dispersion and show the predicted distribution of IMBH masses from the runaway collision model.

\begin{figure*} 
\centering
\includegraphics[scale=0.55]{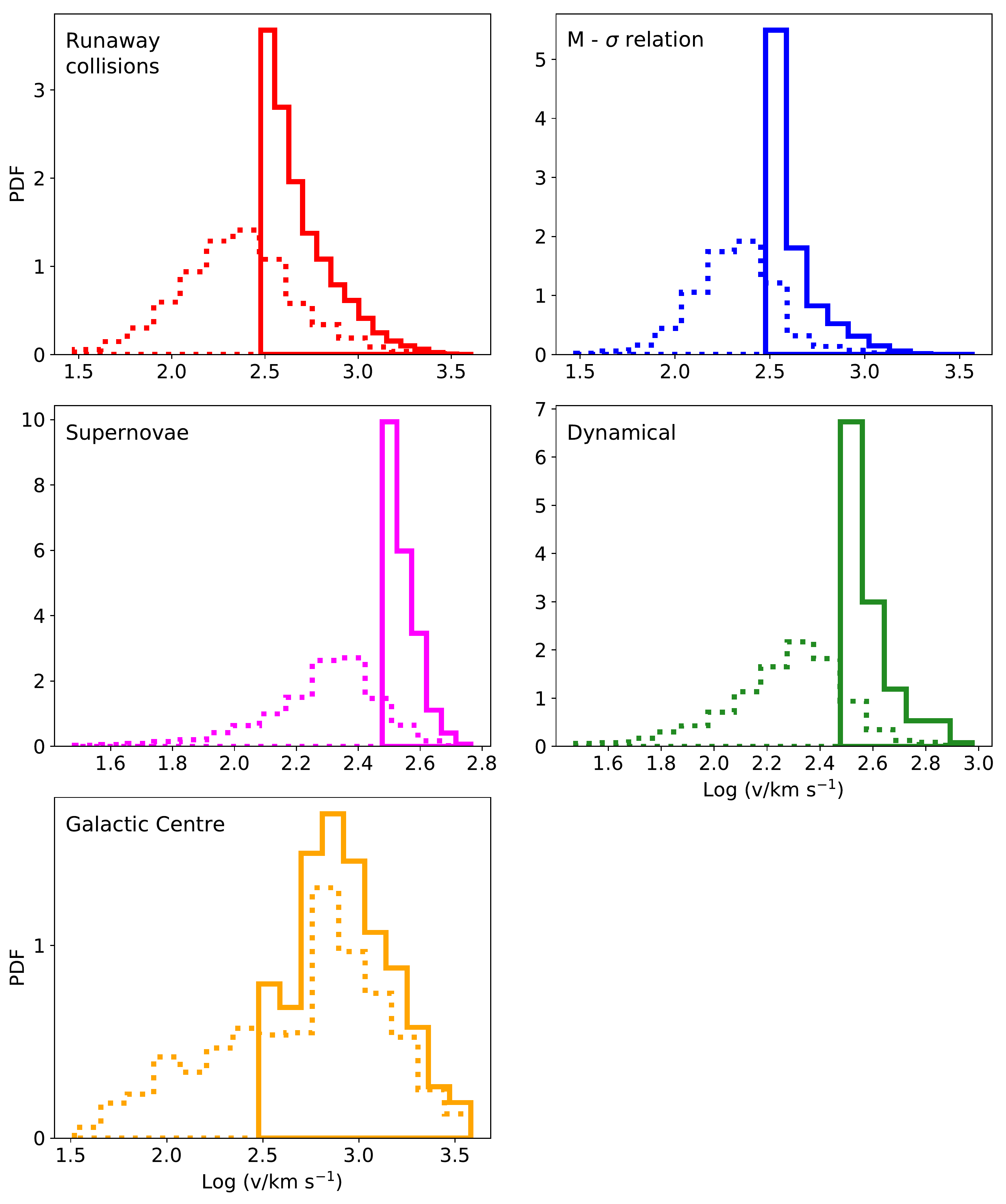}
\caption{Predicted Galactocentric velocity distribution of high-velocity $3\msun$ stars from different formation channels: ejection from an IMBH in the Milky Way's globular clusters (top), dynamical encounters and supernova explosions (middle), Hills mechanism in the Galactic Centre (bottom). Solid lines show the distributions with a cut in the velocity at $30\kms$, dotted lines show the distributions with a cut in the velocity at $300\kms$.}
\label{fig:veldist3}
\end{figure*}

We use the distributions of ejection velocities obtained with the scattering experiments (see Sect. \ref{sect:distributions}) to estimate the statistical properties of high-velocity stars ejected from the Galactic population of globular clusters. Here we only use the population of clusters with a measured value of $\sigma$, in order to make a fair comparison between the two different IMBH models. We then compare these properties with those of HVSs ejected from the Galactic centre as well as RSs ejected from the Galactic disc. To generate a population of HVSs originating in the Galactic Centre, we use the method outlined in \citet{brm06}, that we briefly described in Sect.~\ref{sect:hills}-\ref{sect:clusters} (see Eq.~\ref{eqn:vej}). 
We also create a mock population of RSs following the prescriptions of \citet{ken14}. We assume that the initial space density of RSs follows the space density of stars in the Galactic disc \citep{bov13}
\begin{equation}
f(r_{\rm RS})\propto r_{\rm RS}\ e^{-r_{\rm RS}/r_0}\ ,
\end{equation}
in the range $3$-$30$ kpc, with $r_0=2.4$ kpc \citep{bra07}. 
For the typical velocity distributions of RSs, we distinguish between two cases: supernova ejections and dynamical ejections. In the supernova ejection mechanism \citep{port00,brm09} we model the velocity distribution as
\begin{equation}
f(v_{\rm SN})\propto e^{-v_{\rm SN}/v_{\rm SN,0}}\ ,
\end{equation}
in the range $20\kms$-$400\kms$, with $v_{\rm SN,0}=150\kms$. For the dynamical ejection channel we follow the prescriptions by \citet{persu12} and assume an ejection velocity distribution
\begin{equation}
f(v_{\rm DY})\propto 
\begin{cases}
v_{\rm DY}^{-3/2} &\ v_{\rm DY}\le v_{\rm DY,0} \\
\\
v_{\rm DY}^{-8/3} &\ v_{\rm DY}> v_{\rm DY,0} \\
\end{cases}
\end{equation}
in the range $20-800\kms$, with $v_{\rm DY,0}=150\kms$.

As discussed in Sect.~\ref{sect:clusters}, we follow the procedure outlined in \citet{ken14} to compute the observational properties of the ejected stars, for all the mechanisms considered above. In particular, for each ejected star we randomly choose an ejection time $t_{ej}$ and an observation time $t_{obs}$ between zero and the star's main-sequence lifetime $t_{ms}$. Whenever $t_{ej}<t_{obs}$, we assign the star an ejection velocity according to the distribution appropriate for its origin, and combine it with either the cluster orbital velocity (for IMBH ejections) or the Galactic disc velocity (for dynamical and supernova RSs). Finally, we integrate the full 3D orbit of the star through the Galaxy up to a maximum time $T=t_{obs}-t_{ej}$.

\subsection{Velocity distributions}

\begin{figure} 
\centering
\includegraphics[scale=0.55]{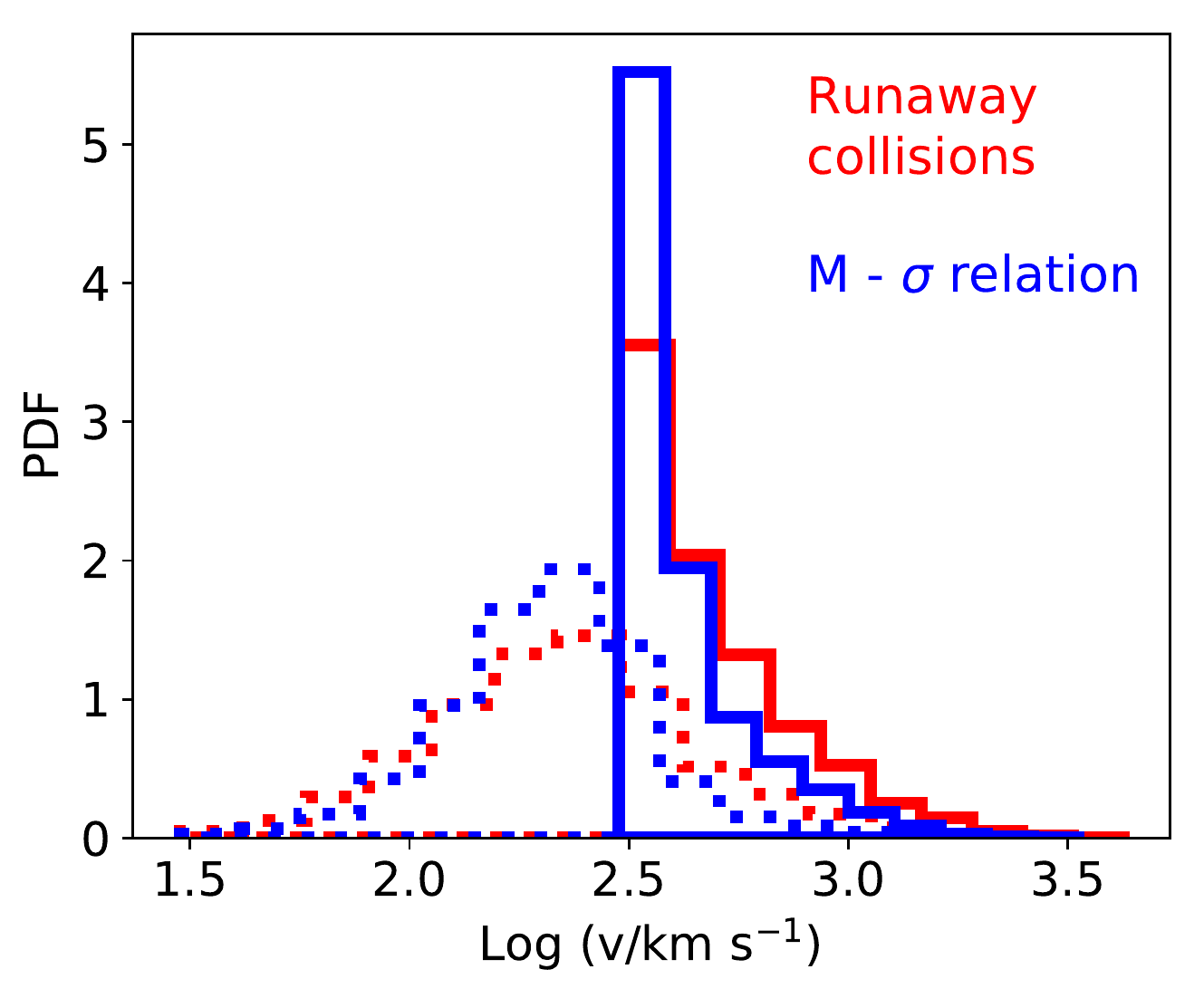}
\caption{Predicted Galactocentric velocity distribution of high-velocity $3\msun$ stars from an IMBH in the Milky Way's globular clusters, using Eq.~\ref{eqn:sigfit} for clusters without a measured value of $\sigma$.}
\label{fig:allsigma}
\end{figure}

In Figure~\ref{fig:veldist3} we compare the distribution of velocities of high-velocity $3\msun$ stars produced by the different channels: ejection by an encounter with an IMBH in a Galactic globular cluster (top panels), ejection from the disc by either supernova explosions in binaries or dynamical encounters (middle panels) and Hills mechanism in the Galactic Centre (bottom). The velocities are taken at the time the star is observed, to allow for a more meaningful comparison with observations of HVSs in the halo. Velocities are cut at either $300\kms$ (solid lines, for comparison with HVSs) or $30\kms$ (dotted lines, for comparison with runaways). We find that the distributions of observed velocities for the IMBH channel are peaked around $300\kms$, with a tail extending to $\sim 2000\kms$. The runaway collision scenario predicts larger velocities than the $M-\sigma$ model. This is justified by the fact that the typical IMBH mass is higher in the former model. We stress that we have only used the population of clusters with a measured value of $\sigma$ for this comparison between the two different IMBH models. In addition, we have derived a scaling relation from Harris catalogue
\begin{equation}
\sigma=8.2\kms \left(\frac{M_{\rm GC}}{10^5\msun}\right)^{0.45}\ ,
\label{eqn:sigfit}
\end{equation}
that we have used to repeat our calculation, now using the whole sample of Galactic globular clusters. We find similar results in this case (see Fig.\,\ref{fig:allsigma}), but we leave a more detailed comparison of the two IMBH mass models for future work, when velocity dispersion data will become available for all the known Galactic globular clusters. The distributions for ejection from the Galactic disc have similar peaks, though they extend only to a maximum of $\sim 800\kms$ for the dynamical ejections and $\sim 400-500\kms$ for supernova ejections. The distributions for Galactic Centre ejections are broader, with a peak at $\sim 100\kms$ and a tail extending up to $\sim 3000\kms$. Differences arise also for lower-mass stars, though they are more pronounced for the more massive stars.

\subsection{Galactocentric distance and Galactic latitude distributions}
\begin{figure*} 
\centering
\includegraphics[scale=0.55]{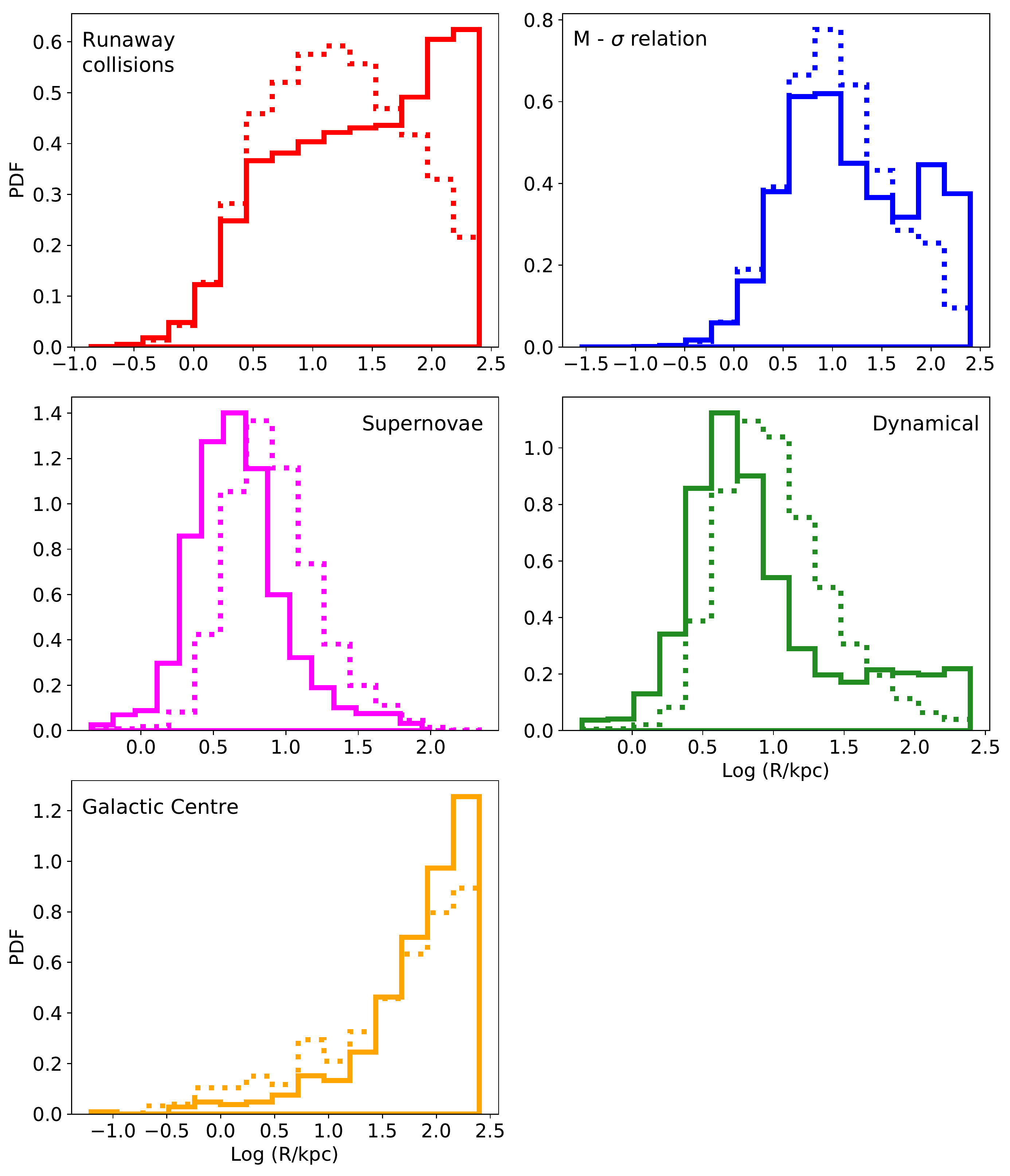}
\caption{Predicted distribution of Galactocentric distances of high-velocity $3\msun$ stars from different formation channels: ejection from an IMBH in the Milky Way's globular clusters (top), dynamical encounters and supernova explosions (middle), Hills mechanism in the Galactic Centre (bottom). Solid lines show the distributions with a cut in the velocity at $30\kms$, dotted lines show the distributions with a cut in the velocity at $300\kms$.}
\label{fig:raddist3}
\end{figure*}
The distributions of Galactocentric distances of high-velocity $3\msun$ stars are shown in Figure~\ref{fig:raddist3} for the different formation channels.  Velocity cuts at $30\kms$ and $300\kms$ are applied in this case as well. The distributions for the different channels are significantly different, though they all peak at distances of a few kpc, except from the Hills mechanism peaked at $\sim 200$ kpc. The disc ejections (dynamical and SN) have a fast decline, with a small tail up to $100$ kpc. The largest distances are reached with Galactic centre ejections.

\begin{figure*} 
\centering
\includegraphics[scale=0.55]{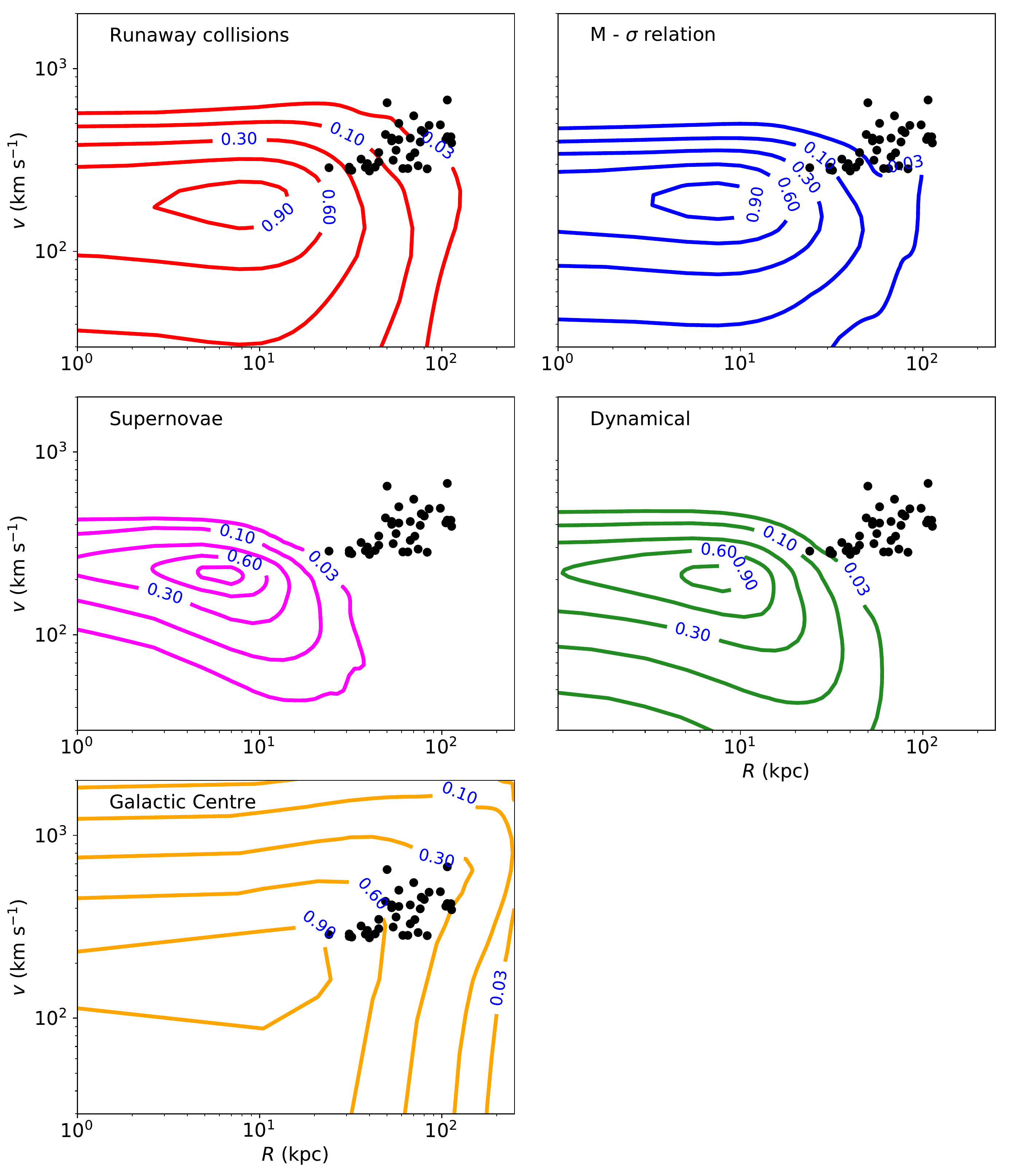}
\caption{Predicted density distribution of high-velocity $3\msun$ stars from different formation channels: ejection from an IMBH in the Milky Way's GCs (top), dynamical encounters and supernova explosions (middle), Hills mechanism in the Galactic Centre (bottom). For comparison, data from the \citet{brw14} sample of HVSs is shown with filled circles.}
\label{fig:map}
\end{figure*}

A comparison with available HVS data \citep{brw14} is given in Fig.\,\ref{fig:map} through a density map in Galactocentric velocity versus distance. Predictions from a Hills-type mechanism provide the best match to current observations, with both ejections from the GC and ejections from a cluster IMBH with a mass extracted from the  $M-\sigma$ relation showing good agreement with data. A more detailed comparison with data taking into account observational biases would be required to draw stronger conclusions but is beyond the scope of this work.

\begin{figure*} 
\centering
\includegraphics[scale=0.55]{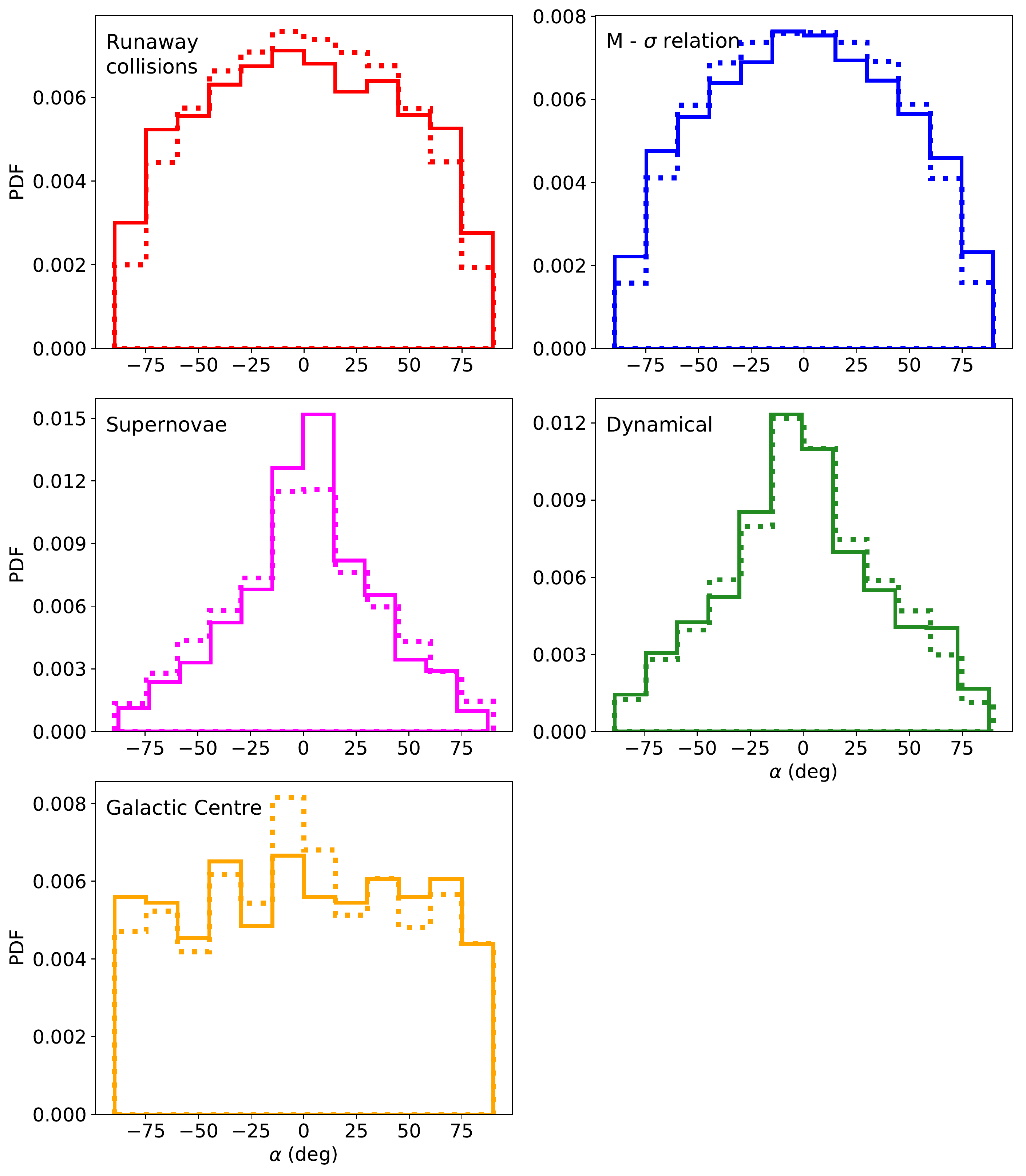}
\caption{Predicted distribution of Galactic latitude of high-velocity $3\msun$ stars from different formation channels: ejection from an IMBH in the Milky Way's globular clusters (top), dynamical encounters and supernova explosions (middle), Hills mechanism in the Galactic Centre (bottom). Solid lines show the distributions with a cut in the velocity at $30\kms$, dotted lines show the distributions with a cut in the velocity at $300\kms$.}
\label{fig:alphadist3}
\end{figure*}
In Figure~\ref{fig:alphadist3} we show  the predicted distributions of the Galactic latitude $\alpha$ for the different models. Runaway stars originated in the Galactic disc are clearly clustered close to the Galactic plane, with a peak at $\alpha \sim 0$ and symmetric distributions around the plane. High-velocity stars coming from clusters hosting IMBHs have a wider distribution of Galactic latitudes, roughly uniform in the range (-75,+75) deg, with fewer stars at $\pm 90$ deg with respect to the Galactic plane.
Stars ejected from the Galactic centre have a uniform distribution in $\alpha$ since they are ejected isotropically by the SMBH. 

As in the case of the velocity distributions, differences arise also for less massive stars, but they are more pronounced for the more massive ones.

\subsection{Rates}
Ejection rates for high-velocity stars are different depending on the ejection mechanism. 
In the case of the dynamical and supernovae ejections, the estimated rate is $\sim 10^{-3}$ yr$^{-1}$, assuming that our Galaxy forms stars at $\sim 1\msun$ yr$^{-1}$ according to a Kroupa initial mass function, and that $0.1\%$ of them undergoes high-velocity ejections \citep{brw15}. 
In the empty loss-cone regime, the Galactic Centre ejects stars through the Hills mechanism with a typical rate of $\sim 10^{-4}$-$10^{-5}$ yr$^{-1}$ \citep{yut03}.

For the IMBH channel, the rate is essentially dominated by the most massive IMBHs. Assuming as a typical IMBH mass of $1000\msun$ and a typical stellar dispersion velocity of $10\kms$, the rate of ejection of high-velocity stars in a given cluster is $\sim 4 f_b \times 10^{-5}$ yr$^{-1}$ (see Eq.\,\ref{eq:imbhrate}), where $f_b$ is the cluster binary fraction. For the overall population of clusters in the galaxy, the total rate is $\sim 4 f_b \zeta N_{\rm cl} \times 10^{-5}$ yr$^{-1}$, where $N_{\rm cl}$ is the number of globular clusters and $\zeta$ is the fraction of clusters that harbour an IMBH. \citet{gie15} found that the fraction of clusters that may host an IMBH can be as high as $20\%$. Assuming a typical binary fraction $f_b=0.1$, this translates into a total rate of $\sim 1.2 \times 10^{-4}$ yr$^{-1}$. We note that the binary fraction generally decreases over time, though typically only to  $\sim 60$-$80\%$ of the initial value, both for shallow Plummer and concentrated King profiles \citep{tren07}. This implies that the  total rate of ejections for the IMBH channel decreases to $\sim 0.7$-$1 \times 10^{-4}$ yr$^{-1}$ in time.

Given the rates above, we can conclude that high-velocity stars ejected from globular clusters harbouring IMBHs should be $\sim 10$ times less likely than runaway stars ejected from the Galactic disc and similarly a factor of a few more likely than Hills ejections from the Galactic centre, though we caution that uncertainties in the rates remain significant.

\begin{figure} 
\centering
\includegraphics[scale=0.55]{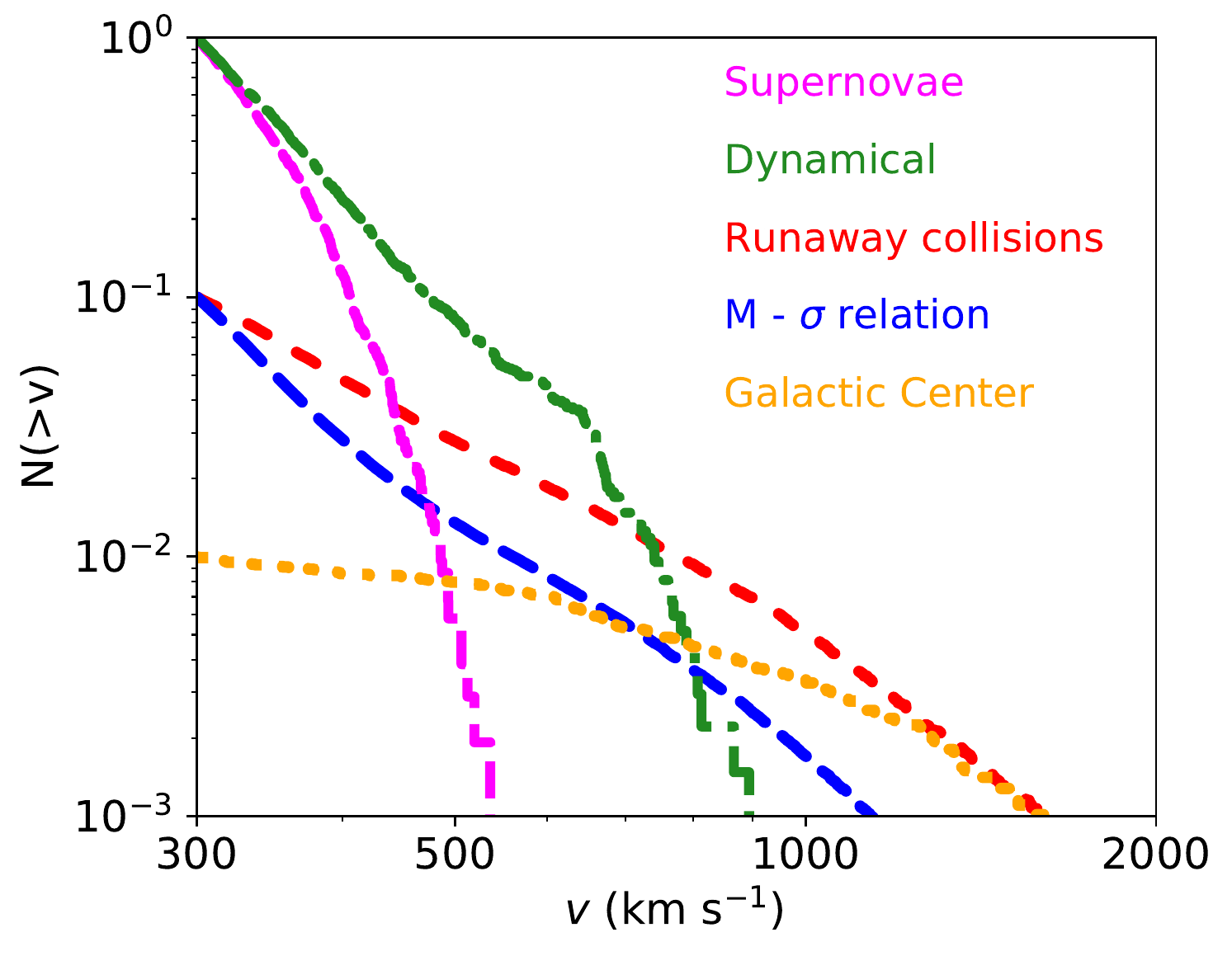}
\caption{Cumulative distributions of observed velocities of high-velocity stars from the different channels, each cut at $300\kms$ and normalised to the predicted rate of the channel.}
\label{fig:normvel}
\end{figure}
In Fig.\,\ref{fig:normvel} we show the cumulative distributions of observed velocities for high-velocity stars (within $100$ kpc) from the different channels, normalised to the supernova/dynamical scenario rate. As expected, the population of high-velocity stars is dominated by runaway stars from the Galactic disc at low velocities ($V \lesssim 500\kms$) and high-velocity stars from IMBH ejections from clusters at high velocities ($V \gtrsim 500\kms$), with HVSs from the GC providing a significant contribution to the high-velocity tail ($V \gtrsim 1000\kms$).

\section{Discussions and Conclusions}
\label{sect:concl}

The \textit{Gaia} mission has revolutionized astrometry and is offering an unprecedented opportunity to study the population of high-velocity stars in our Galaxy, thanks to its precise proper motion measurements. The full \textit{Gaia} data release will allow to determine the origin of (most) HVSs and put constraints on the Galactic mass distribution and dark halo potential \citep{gnedin2010,fl2017,rossi2017}.

The leading scenario to explain the velocities of the hypervelocity stars in the halo is the classical Hills mechanism \citep{hills88}, where a stellar binary is unbound by the SMBH in the Galactic centre, as a result of which one star is captured in orbit around the SMBH while the companion is ejected with a typical velocity of a thousand $\kms$. This model predicts HVSs moving away from the Galactic centre on nearly radial orbits.  Surprisingly, recent observations show that only a few of the known HVSs (the ones with the highest velocities) can be traced back to the centre of the Galaxy \citep{boub18,brown18,march18, erk19}. \citet{keny18} showed that the Galactic disc and the LMC potential may play a role in deflecting HVSs from a nearly radial orbit. Yet, the origin of most of the known HVSs remains unknown.

We considered the ejection of high velocity stars from star clusters hosting an IMBH due to encounters between binary stars and the IMBH itself. We performed a large number of high-precision scattering experiments varying the IMBH mass and the properties of the binaries and derived the velocity and space distributions of the ejected stars by taking into account the effect of the parent cluster's orbital motion. We found that the properties of the ejected stars, which can contribute both to the Galactic RS and HVS populations, depend on the cluster orbit, with most of the unbound HVSs found at Galactocentric distances $R\gtrsim 30$ kpc. 

As expected, HVSs generated by a Hills-like mechanism in a star cluster hosting an IMBH would not travel on orbits consistent with a Galactic centre origin, rather they would point back to their original host cluster, thus providing observational evidence for the presence of an IMBH. We caution, however, that observational errors in the distance, radial velocity and proper motions would propagate while backtracing the orbit in the Galactic potential, thus making a precise localisation of the birth location in the disc challenging. In addition to a population of unbound HVSs, this mechanism can produce a large population of bound RSs \citep*{silva11,zhong14,vickers15,subr2019}.

We also modelled the ejection of high-velocity stars from the Galactic population of globular clusters, assuming that each cluster contains an IMBH and combining the ejection velocity of the star with the cluster's orbital velocity, as well as propagation of the star in the Galactic potential up to the estimated detection time. We compared the properties of the resulting high-velocity stars with those of mock populations ejected from both the Galactic disc (dynamical and SN ejections) and the Galactic centre (Hills mechanism). We find that high-velocity stars ejected by IMBHs have distinctive distributions in velocity, Galactocentric distance and Galactic latitude. The ejection rate is $\sim 0.7-1.2\times10^{-4}$ yr$^{-1}$, similar or slightly larger than the rate for the Hills mechanism. Altogether, this opens up the possibility of constraining the origin of high-velocity stars with future observations. A detailed comparison with available data requires a detailed analysis of observational selections and biases and is left to future work.

Studying spectroscopic and kinematic data of runaways and HVSs can help constraining their origin and the possible presence of an IMBH in the core of a given globular cluster. Spectroscopic data can be used to obtain radial velocities, while tangential velocities can be obtained from proper motion measurements, whose combination, along with the sky position of the stars, gives the full 6-D phase space information to determine their trajectories. This approach has already proven successful in a few cases in which high-velocity stars were traced back to their birth location, providing indirect evidence for the existence of IMBHs in certain clusters. \citet{hoo01} used radio observations and milli-arcsecond accuracy astrometry from the \textit{Hipparcos} satellite of the orbits of $56$ RSs and nine compact objects with distances $\lesssim 700$ pc to identify their parent stellar group. \citet{hebe08} studied the mass, evolutionary lifetime and kinematics of HD 271791 by using proper motion measurements from a collection of catalogues, and found that the likely birthplace of HD 271791 is the outer Galactic disc, while the Galactic Centre origin is ruled out. Recently, \citet{lenno18} and \citet{renz18} used \textit{Gaia} data to show that the dynamics of the very-massive runaways VFTS 16 and VFTS682 are consistent with ejection from the young massive cluster R136. \citet{hatt18} studied the origin of hyper-runaway subgiant LAMOST-HVS1 by using \textit{Gaia} data, and found that it was likely ejected dynamically from near the Norma spiral arm as a consequence of a few-body encounter or a Hills-like ejection by an IMBH. Thanks to present and future high precision data from \textit{Gaia}, similar studies can be performed for known and candidate high-velocity stars, which might have been ejected through the mechanism discussed in this work, thus revealing new IMBH candidates.

\section{Acknowledgements}
We thank the referee for insightful comments, which were very valuable for improving the paper. GF is supported by the Foreign Postdoctoral Fellowship Program of the Israel Academy of Sciences and Humanities. GF also acknowledges support from an Arskin postdoctoral fellowship at the Hebrew University of Jerusalem. GF thanks Seppo Mikkola for helpful discussions on the use of the code \textsc{archain}. Simulations were run on the \textit{Astric} cluster at the Hebrew University of Jerusalem.

\bibliographystyle{mn2e}
\bibliography{biblio}
\end{document}